\documentclass[review,1p,number]{elsarticle}

\usepackage[utf8]{inputenc}
\usepackage[T1]{fontenc}
\usepackage{lmodern}
\usepackage{anyfontsize}
\usepackage[section]{placeins} 
\usepackage{stmaryrd} 
\SetSymbolFont{stmry}{bold}{U}{stmry}{m}{n} 

\usepackage{graphics}
\usepackage{ifthen}
\usepackage{enumerate}

\usepackage{savesym}
\usepackage{amsmath}
\savesymbol{Bbbk}
\usepackage{amssymb} 
\restoresymbol{AMSSYMB}{Bbbk}

\RequirePackage{amsthm}		
\RequirePackage{amsfonts}	
\RequirePackage{bbm} 		
\RequirePackage{verbatim}	
\RequirePackage{mathrsfs}	
\RequirePackage{dsfont}

\usepackage{subcaption}

\usepackage{tikz}
\usetikzlibrary{calc}

\usepackage{appendix}

\input{commandes_perso.tex}

\usepackage{hyperref}

\renewcommand{\emptyset}{\varnothing}
\newcommand{\uniondisjointe}{+}
\newcommand{\fulljoin}{\oplus}

\newtheorem{theorem}{Theorem}
\newtheorem{lemma}[theorem]{Lemma}
\newtheorem{coro}[theorem]{Corollary}
\newtheorem{redrule}{Reduction rule}
\newtheorem{prop}[theorem]{Proposition}
\newtheorem{claim}[theorem]{Claim}

\title{A quasi-quadratic vertex Kernel for Cograph edge editing}

\author[1]{Christophe Crespelle%
\fnref{fn1}}
\ead{christophe.crespelle@univ-cotedazur.fr}

\author[2]{Rémi Pellerin\corref{cor1}%
\fnref{fn2}}
\ead{remi.pellerin@ens-lyon.fr}

\author[2]{Stéphan Thomassé%
\fnref{fn3}}
\ead{stephan.thomasse@ens-lyon.fr}

\affiliation[1]{organization={Université Côte d'Azur, Laboratoire
	d’Informatique, Signaux et Systèmes de Sophia-Antipolis (I3S)},
addressline={2000, route des Lucioles - Les Algorithmes - bât. Euclide B},
postcode={06900},
city={Sophia Antipolis},
country={France}}

\affiliation[2]{organization={Univ Lyon, EnsL, UCBL, CNRS, Inria,  LIP, F-69342},
addressline={46, allée d’Italie},
postcode={69364},
city={Lyon cedex 07},
country={France}}

\cortext[cor1]{Corresponding author}
\fntext[fn1]{christophe.crespelle@univ-cotedazur.fr}
\fntext[fn2]{remi.pellerin@ens-lyon.fr}
\fntext[fn3]{stephan.thomasse@ens-lyon.fr}

\begin{document}

\begin{abstract} 
	We provide a $O(k^{2}\log k)$ vertex kernel for cograph edge editing. This
	improves a cubic kernel found by Guillemot, Havet, Paul and Perez
	\cite{GHPP2013} which involved four reduction rules. We generalize one of
	their rules, based on packing of induced paths of length four, by
	introducing $t$-modules, which are modules up to $t$ edge modifications. The
	key fact is that large $t$-modules cannot be edited more than $t$ times, and
	this allows to obtain a near quadratic kernel. The extra $\log k$ factor
	seems tricky to remove as it is necessary in the combinatorial lemma on
	trees which is central in our proof. Nevertheless, we think that a quadratic
	bound should be reachable.
\end{abstract}


\maketitle

\section{Introduction}
A particularly large class of graph algorithmic questions can be seen as
modification problems. Such problems are defined by a target class of graphs
$\mathcal C$ and the types of modifications allowed on a graph, such as vertex
deletion or edge addition for example. The question is, given an input graph
$G$, to find the minimum number of such modifications to be performed on $G$ in
order to obtain a graph $H\in\mathcal C$. For instance, the very popular
vertex-cover problem can be seen as a vertex deletion problem in which one wants
to reach the class of edgeless graphs. Also, the feedback-vertex-set problem can
be seen as vertex deletion toward the class of forests.

In these two examples, allowing vertex additions would not make sense as adding
vertices would not help to reach the target class. The situation is the same for
all hereditary target classes, i.e. classes closed by induced subgraphs, which
turns out to be a property shared by the vast majority of the target classes
considered in modification problems (see \cite{Man08} for example). For the case
of edge modification problems, which we consider here, the situation is quite
different as both deletion and addition of edges may help in order to reach some
hereditary target class. Consequently, three kinds of edge modification problems
are classically considered: the deletion problem, in which only deletion of
edges is allowed, the completion problem, allowing only addition of edges and
the editing problem, where both addition and deletion are allowed. The question
asked by edge modification problems is very natural in the sense that one can
assume that the input graph $G$ is a noisy version of a graph $H$ of $\mathcal
C$ in which a small set $S$ of $k$ pairs of vertices has been
modified~\cite{Natanzon1999ComplexityCO}. This is the reason why several edge
modification problems are successfully used in practice to analyse real-world
datasets. As an example of this success, the community detection problem, which
is a central topic in complex networks analysis, is formalised by the
\emph{cluster editing} problem~\cite{Shamir2002ClusterGM}, which asks whether it
is possible to edit at most $k$ pairs of vertices to make the input graph a
disjoint union of cliques, also known as \emph{cluster graphs}.

Unfortunately, most edge modification problems, including \emph{cluster
editing}, are $NP$-hard, even if the target class is very
simple~\cite{Natanzon1999ComplexityCO}. The most striking example of this is the
$NP$-hardness of the editing problem toward the class of graphs that are the
disjoint union of a single clique and an independent set, called \emph{clique +
independent set}. In order to deal with this difficulty of computation, edge
modification problems have often been studied in the framework of parameterized
complexity, see~\cite{CDF+20} for a survey on the topic. In this framework, the
complexity one wants to reach is $f(k)n^c$, where $k$ is the maximum number of
edits allowed in the decision problem, and not the obvious $O(n^k)$ one can
obtain by brute force. A common technique to design such algorithms, called FPT
(for \emph{Fixed Parameter Tractable}), is kernelization. A kernel is a
preprocessing algorithm aiming at reducing in polynomial time (in $n$) the
instance of a problem to an equivalent instance of size bounded by $f(k)$. Such
a kernel is said to be polynomial whenever its size $f(k)$ is (at most)
polynomial in $k$. It is well-known that a problem is FPT if and only if it has
a kernel~\cite{downey1999parameterized}, but not all FPT problems admit a kernel
of polynomial size~\cite{BodlaenderDFH09} (under some complexity hypothesis).
The research for compact kernels for edge modification problems is very
flourishing~\cite{CDF+20} and has achieved remarkable results. For example,
there exists a $2k$ vertex kernel for cluster editing \cite{CaoC12,ChenM12} and
very recently, \cite{DBLP:journals/corr/abs-2105-09566} designed a sublinear
vertex kernel for edge deletion to clique + independent set, which is the first
and, up to this day only, sublinear vertex kernel for an edge modification
problem.



Here, we aim at designing a kernel for the editing problem toward the class of
cographs, which is a proper and natural generalisation of the two classes
mentioned above. Indeed, cographs are the graphs obtained from single vertices
under the closure of two operations: the disjoint union of graphs and their
complete union\footnote{The complete union of two graphs $G_1$ and $G_2$ is
their disjoint union plus all the possible edges between $G_1$ and $G_2$}.
Equivalently, they can also be defined as the graphs with no induced $P_4$ (path
on four vertices). Then, the purpose of the editing problem is that no induced
path on four vertices can be found in the edited graph $H$. Cographs have
received a huge amount of attention in algorithmic graph theory and have been
shown to admit very efficient solutions to various problems. Related to our
concern here, \cite{GHPP2013} shows that all the three edge modification
problems toward the class of cographs admit a cubic vertex kernel. This kernel
size may still appear a bit large compared to the linear and sublinear vertex
kernels mentioned above for two subclasses of cographs, but the solution
proposed in~\cite{GHPP2013} to reach this cubic size is actually already far
from being obvious. Nevertheless, there may still be some room for improvement
as it seems that the cubic size instances provided in~\cite{GHPP2013} in which
none of the reduction rules apply can be reduced further. This is the goal of
this paper. Our hope is that a finer analysis of this (rather simple) problem
could provide some new reduction rules, maybe useful for other classes. Our main
idea is to provide tools in order to roughly localize where edits should
happens. More precisely, we provide upper bounds on the number of edits
performed across a cut $(X,V\setminus X)$. For this, we relax the notion of
module to some approximate version ($t$-module), and argue that not too many
edits can cross a $t$-module. One very nice property of the resulting reduction
rule is that it does not depend on the parameter $k$. This means that the rule
can apply independently of the possibly large value of $k$, which is crucial in
practice to reduce difficult instances.

\section{Notations}

\begin{figure}[th]
	\centering
	\begin{subfigure}{0.45\textwidth}
		\centering
		\scalebox{0.55}{\begin{tikzpicture}[line width=0.5mm,scale=2]
	\node[draw,fill,circle] (0) at (0,2) {};
	\node[draw,fill,circle] (1) at (0,1) {};
	\node[draw,fill,circle] (4) at (0,-1) {};
	\node[draw,fill,circle] (5) at (0,-2) {};

	\node[draw,fill,circle] (2) at (-2,0) {};
	\node[draw,fill,circle] (3) at (2,0) {};

	\draw[-,dashed,red] (0) to (1);
	\draw[-] (1) to (3);
	\draw[-] (3) to (5);
	\draw[-] (5) to (4);
	\draw[-,dashed,red] (4) to (2);

	\draw[-] (0) to (3);
	\draw[-,dashed,red] (1) to (2);
	\draw[-] (2) to (5);
	\draw[-] (3) to (4);
\end{tikzpicture}}
		\caption{not a cograph (induced $P_4$ in dotted red)}
	\end{subfigure}
	\hfill
	\begin{subfigure}{0.45\textwidth}
		\centering
		\scalebox{0.55}{\begin{tikzpicture}[line width=0.5mm,scale=2]
	\node[draw,fill,circle] (0) at (0,2) {};
	\node[draw,fill,circle] (1) at (0,1) {};
	\node[draw,fill,circle] (4) at (0,-1) {};
	\node[draw,fill,circle] (5) at (0,-2) {};

	\node[draw,fill,circle] (2) at (-2,0) {};
	\node[draw,fill,circle] (3) at (2,0) {};

	\draw[-] (0) to (1) to (3) to (5) to (4) to (2) to (0);
	\draw[-] (0) to (3);
	\draw[-] (1) to (2);
	\draw[-] (2) to (5);
	\draw[-] (3) to (4);
\end{tikzpicture}}
		\caption{a cograph}
	\end{subfigure}
	\caption{Example of a cograph and of a non cograph}
	\label{fig:cograph_and_not_cograph}
\end{figure}
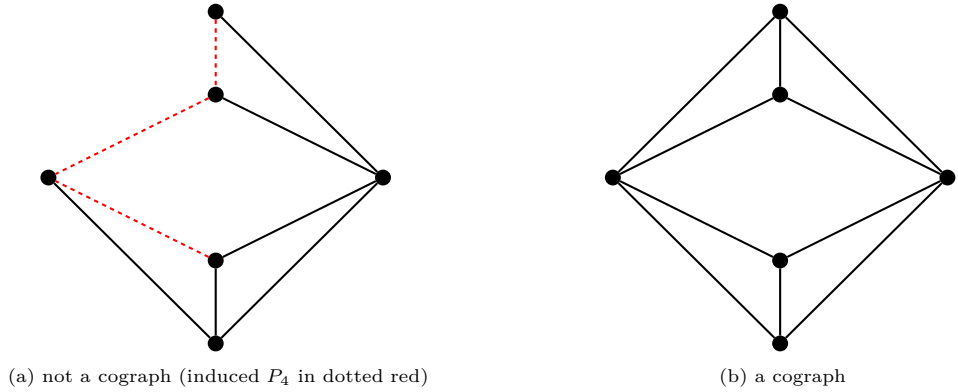

We denote by $P_4$ the path on four vertices. A \emph{cograph} is a graph which
does not contain any induced $P_4$. Figure \ref{fig:cograph_and_not_cograph}
shows an example of a cograph and an example of a non cograph with an induced
$P_4$ in dotted red. Let $H=(V,E)$ be a cograph and $S$ be a subset of pairs of
vertices of $H$. We call \emph{edit of~$H$ by $S$} the graph $G$ obtained from
$H$ by changing the adjacency relation of the elements of $S$, i.e.~$G$ differs
from $H$ for every pair of vertices in $S$ and coincides for the pairs not in
$S$. More formally, $G=(V,E \vartriangle S)$. Since all the graphs that we will
consider are simple graphs, such a set $S$ will always satisfies that $(x,y) \in
S$ if and only if $(y,x) \in S$. The general editing problem for a fixed
class $\mathcal C$ of graphs is, given an input graph $G$ and an integer $k$, to
ask for the existence of an edit $H$ of $G$ by some set of pairs $S$ of size at
most $k$ such that $H\in \mathcal C$. This is the parameterized version of
\emph{$\mathcal C$-editing problem}.

Observe that $H$ is an edit of $G$ by $S$ whenever $G$ is an edit of $H$ by $S$. Taking the
opposite point of view will be useful as we understand better the structure of $H$ since it is a
cograph. Though, all along this paper $H$ is a cograph on vertex set $V$ and
$G$ is an edge editing of $H$ by
a set $S$ of pairs of vertices of size at most $k$.  Given a subset $X$ of vertices, we call
\emph{$X$-cut} the set $\delta (X)$ of pairs of vertices $xy$ where $x\in X$ and $y\notin X$. The set
of neighbors of a vertex $x$ is denoted by $N(x)$. When $X$ is a subset of vertices of a graph $G$, we
denote by $G[X]$ the subgraph induced by $G$ on $X$. 

\begin{figure}[t]
	\centering
	\begin{subfigure}{0.4\textwidth}
		\centering
		\scalebox{0.475}{\begin{tikzpicture}[line width=0.5mm,scale=2]
	\node[draw,circle] (0) at (0,2) {$0$};
	\node[draw,circle] (1) at (0,1) {$1$};
	\node[draw,circle] (4) at (0,-1) {$4$};
	\node[draw,circle] (5) at (0,-2) {$5$};

	\node[draw,circle] (2) at (-2,0) {$2$};
	\node[draw,circle] (3) at (2,0) {$3$};

	\draw[-] (0) to (1) to (3) to (5) to (4) to (2) to (0);
	\draw[-] (0) to (3);
	\draw[-] (1) to (2);
	\draw[-] (2) to (5);
	\draw[-] (3) to (4);
\end{tikzpicture}}
	\end{subfigure}
	\hfill
	\begin{subfigure}{0.4\textwidth}
		\centering
		\begin{tikzpicture}
	\node{$\fulljoin$} [sibling distance = 2cm]
		child{node{$\uniondisjointe$} [sibling distance = 1cm]
					child{node{$2$}}
					child{node{$3$}}
				}
		child{node{$\uniondisjointe$} [sibling distance = 2cm]
					child{node{$\fulljoin$} [sibling distance = 1cm]
								child{node{$0$}}
								child{node{$1$}}
							}
					child{node{$\fulljoin$} [sibling distance = 1cm]
								child{node{$4$}}
								child{node{$5$}}
							}
				};
\end{tikzpicture}
	\end{subfigure}
	\caption{A cograph and its cotree}
	\label{fig:decomposition_tree}
\end{figure}
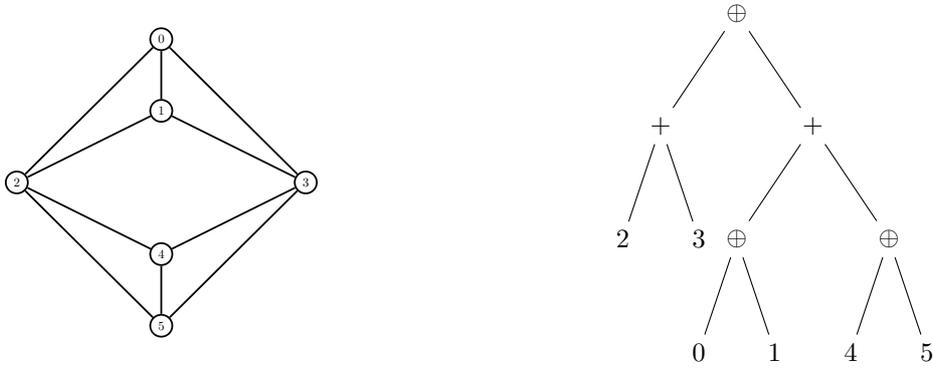

The most useful characterization of cographs is their \emph{cotree}. Precisely,
for any
cograph $H$, there exists a rooted tree $T$ whose leaves are identified to the vertices of $H$ and whose
internal vertices have at least two children and are labelled by
$\uniondisjointe$ or $\fulljoin$. Moreover, two vertices $x,y$ form an edge of
$H$ if and only if their closest ancestor is labelled $\fulljoin$. A proof of this result can be found
in~\cite{CORNEIL1981163}. There are several possible choices for this
tree $T$, but there is a canonical one if every child of a node labelled
$\uniondisjointe$ has label $\fulljoin$ and every
child of a node labelled $\fulljoin$ has label $\uniondisjointe$ (see
\cite{CORNEIL1981163}). For instance, the cotree of a clique has a unique
internal node labelled $\fulljoin$. Another cotree for a less
specific example is shown on Figure~\ref{fig:decomposition_tree}.

\section{Reduction rules}

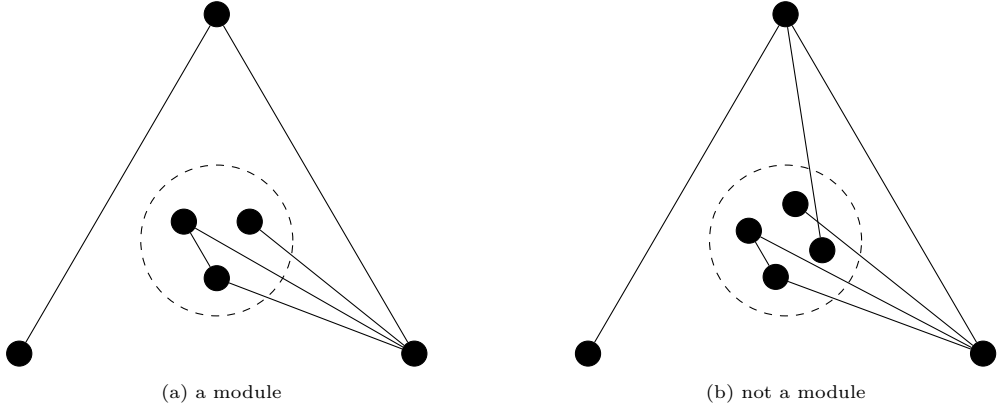
\begin{figure}[t]
	\centering
	\begin{subfigure}{0.45\textwidth}
		\centering
		\scalebox{1}{\begin{tikzpicture}
	\node[draw,fill,circle] (0) at (-30:3) {};
	\node[draw,fill,circle] (1) at (90:3) {};
	\node[draw,fill,circle] (2) at (210:3) {};

	\node[draw,fill,circle] (3) at (30:0.5) {};
	\node[draw,fill,circle] (4) at (150:0.5) {};
	\node[draw,fill,circle] (5) at (-90:0.5) {};

	\draw[-] (0) to (1) to (2);

	\draw[-] (3) to (0);
	\draw[-] (4) to (0);
	\draw[-] (5) to (0);

	\draw[-] (4) to (5);

	\draw[dashed] (0,0) circle (1);
\end{tikzpicture}}
		\caption{a module}
		\label{fig:module}
	\end{subfigure}
	\hfill
	\begin{subfigure}{0.45\textwidth}
		\centering
		\scalebox{1}{\newcommand{\rotation}{-15}
\begin{tikzpicture}
	\node[draw,fill,circle] (0) at (-30:3) {};
	\node[draw,fill,circle] (1) at (90:3) {};
	\node[draw,fill,circle] (2) at (210:3) {};

	\node[draw,fill,circle] (3) at (90+\rotation:0.5) {};
	\node[draw,fill,circle] (4) at (180+\rotation:0.5) {};
	\node[draw,fill,circle] (5) at (270+\rotation:0.5) {};
	\node[draw,fill,circle] (6) at (0+\rotation:0.5) {};

	\draw[-] (0) to (1) to (2);

	\draw[-] (3) to (0);
	\draw[-] (4) to (0);
	\draw[-] (5) to (0);
	\draw[-] (6) to (1);

	\draw[-] (4) to (5);

	\draw[dashed] (0,0) circle (1);
\end{tikzpicture}}
		\caption{not a module}
		\label{fig:2module}
	\end{subfigure}
	\caption{Example and counter example of modules}
	\label{fig:module_and_not_module}
\end{figure}

In \cite{GHPP2013}, the authors show that the cograph editing problem has a cubic kernel. Their reduction rules are
mainly based on two features: the induced copies of $P_4$ in $G$, and the modules of $G$. A
\emph{module} is a set of vertices $X$ such that all vertices in $X$ have the same neighborhood in
$V\setminus X$. We say that two vertices are \emph{twins} if they form a module. Figure~\ref{fig:module} gives an example of a module in some graph and Figure~\ref{fig:2module}
a counter example. In this counter example, observe that we can make the set $X$
inside the dotted circle a
module by editing $2$ edges. We say that $X$ is a $2$-module since it is a module up to (at most) $2$
edge edits. In order to define our new reduction rule, will need this notion of $t$-module.

\begin{lemma}
	\label{lem:module_p4}
	Let $G = (V,E)$ be a graph and $X \subseteq V$ be a module of $G$. An induced $P_4$

	\begin{itemize}
		\item either is included in $X$
		\item or is included in $V \setminus X$
		\item or has exactly one vertex in $X$
	\end{itemize}
\end{lemma}

\begin{proof}
	Let $P$ be a $P_4$ that is an induced subgraph of $G$. Observe that $P \cap X$ is a module of $P$. The
	modules of $P$ are the empty set, singletons and $P$ itself which proves the Lemma.
\end{proof}

The crucial fact shown in \cite{GHPP2013} is that for every module $X$ in $G$, one can assume that~$X$
remains a module in a minimum cograph edit $H$. Here is a sketch of the argument. Assume that $S$ is
a minimum cograph set of edits of $G$ so that $H = (V(G), E(G) \vartriangle S)$ and $X$ is a module of $G$. We consider a vertex $x \in X$ which is
incident to the least number of pairs in $S \cap \delta(X)$. We now modify $S$
to $S'$ in such a way that all
vertices in $X$ have the same neighborhood as $x$ in $V \setminus X$. The new
graph $G' = (V(G), E(G) \vartriangle S')$ that we obtain has no $P_4$
since the only copy $C$ of some $P_4$ we could have created by modifying $S$ intersects both $X$ and
$V \setminus X$. But $X$ is a module of $G'$, so $C$ has only one vertex in $X$ by
Lemma~\ref{lem:module_p4}, for instance $x$,
which is impossible since $C$ would be an induced $P_4$ in $H$. Therefore this new edition has at most
as many edited pairs as $S$ and leaves $X$ a module. In particular, if $X$ has size more than $k$,
there is no edited pair in $\delta (X)$.

We are now ready to recall the three reduction rules of \cite{GHPP2013} to apply
to $(G,k)$. We slightly reformulated these rules for our needs. In particular,
we use the notion of comodule which proves to be convenient for writing our
proof. A module $M$ is a \emph{comodule} if $M$ is a connected component or is a
connected component in the complement of the graph. This is equivalent to the
notion of maximal strong module. Despite the fact that our three first rules are a
bit different from these of Guillemot et al, there are actually equivalent in
the following sens: a graph $G$ is reduced under our three first rules if and
only if it is reduced for the three first rules of Guillemot and al. A proof of
this fact can be found in Proposition~\ref{prop:rulesequivalence}.

\begin{redrule} {\rm (comodule rule)}\label{componentrule}
If $G$ has a comodule $C$ which induces a cograph, remove $C$.
\end{redrule}

The safeness of this rule is clear since if we do not edit any pair incident to $C$, no $P_4$ can intersect $C$.

\begin{redrule} {\rm (module reduction rule)}\label{eliminationrule}
	If $G$ has a module $M$ of size $\abs{V(M)} > k+1$ inducing an independent set, reduce $M$ to size $k+1$.
\end{redrule}

This rule is also safe since we can assume that $M$ remains a module, and since
its size is at least $k+1$, no pair of $\delta(M)$ can be edited. 

\begin{redrule} {\rm (module extraction rule)}\label{modulerule}
If $X$ is a module of $G$ which is not a comodule and such that $G[X]$ contains an edge, add a disjoint copy of $G[X]$ to $G$ (no edge between them) and replace the original $G[X]$ by an independent set of size $|X|$.
\end{redrule}

This is a very clever rule since it adds new vertices to $G$, which is precisely
the opposite idea of kernelization! To understand its safeness, observe that
either $G[X]$ is a cograph and it will be removed by Rule~\ref{componentrule}
and therefore the total number of edges in $G$ decreases, or $G[X]$ is not a
cograph but then the cotree (see~\cite{CORNEIL1981163}) has been simplified
since we \ofg{pushed $G[X]$ to its root}.

After applying these three rules until none of them apply, the only modules of
$G$ which are not independent sets are comodules. Hence we will always assume
that our input $(G,k)$ is reduced under these rules before applying our new
reduction rule. 

The cubic kernel in~\cite{GHPP2013} is obtained by adding a last rule: If $G$
has $k+1$ induced copies of $P_4$ pairwise intersecting on vertices $x,y$, then
edit $xy$ and decrease $k$ by 1. This rule is clearly safe since if $xy$ is not
edited, some $P_4$ will survive. However, the fact that this rule is really
different in nature from the others three leaves too much slack, and results in
the cubic bound. The key is to be able to deduce that $xy$ must be edited, even
though we only have $\ell+1$ copies of $P_4$ where $\ell$ is smaller than $k$.
We need for this to be able to say that fewer editions than $k$ are permitted in
some zone of the graph $G$. Unsurprisingly, this can be achieved via a
relaxation of the notion of module.

\section{The fourth rule: budget and t-modules}

The key here is to introduce some control on how many editions can be done
across a cut. The \emph{budget} of a set $X$ of $G$ is the minimum $b$ such that
all minimum cograph edits $S$ of $G$ satisfy $|S \cap \delta(X)| \leq b$.

A \emph{$t$-module} in $G$ is a set of vertices $X$ of $G$ such that by editing
a set $T$ of at most $t$ pairs in $G$, we obtain $G'$ in which $X$ is a module.
We usually assume that $T$ is minimal for this property, in particular $T$ is
included in $\delta(X)$. Figure~\ref{fig:2module} shows a $2$-module inside the
dots. 

\begin{lemma}
	\label{lem:module_budget}
	Let $X$ be a $t$-module such that $\abs{X} > k+t$. If there exists an edge
	editing of size $k$, then the budget of $X$ is at most $t$.
\end{lemma}

\begin{proof}
Assume that there is a cograph editing of $G$ by $T \subseteq \delta (X)$ with size at most $t$ in which $X$
is a module. Assume also that $H$ is a minimum cograph editing of $G$ by $S$ with size at most $k$.
Since $\abs{(S \cup T) \cap \delta(X)} \leq \abs{S} + \abs{T} \leq k + t$ and $\abs{X} \geq k+t+1$, there exists a vertex $x
\in X$ which is not incident to any pair in $(S\cup T) \cap \delta (X)$. Consider now the set $S' := T
\cup (S \setminus \delta(X))$ and denote by $G'$ the edition of $G$ by $S'$. Observe that $X$ is a
module of $G'$. Indeed, all vertices of $X$ have the same neighborhood in $V \setminus X$ since they
coincide with the one of $x$. Hence, by Lemma~\ref{lem:module_p4}, the only copies of $P_4$ which
intersects $\delta(X)$ have
exactly one vertex in $X$ but this is impossible since there would be a $P_4$ in $H$ using $x$. Indeed,
$\delta(x)$ is the same in $H$ and $G'$ by hypothesis. So
$G'$ is a cograph, and thus $\abs{S'} \geq \abs{S}$ so $t = \abs{T} \geq \abs{S \cap \delta(X)}$ which
proves that the budget of $X$ is at most $t$.
\end{proof}

Note that testing if a set $X$ is a $t$-module with size at least $k+t+1$ can be
done in polynomial time since we can first guess the vertex $x\in X$ which is
not incident to the edited edges, and then check if making $X$ a module with the
same neighborhood as $x$ in $V\setminus X$ involves at most $t$ edits.

We now turn Lemma~\ref{lem:module_budget} into a reduction rule. A \emph{nested $t$-module} of $G$ is
a partition of its vertex set into five nonempty pairwise disjoint sets $A,B,C,K,I$ such that:
\begin{itemize}
\item The three sets $A$, $A\cup B$ and $A\cup B\cup C$ are $t$-modules and $A$
	has size $\abs{A} > k+t$.
\item The set $B_\fulljoin$ is the subset of $B$ which is completely joined to $A$ and to
	$K$ and such that there is no edge between $I$ and $B_\fulljoin$.
\item The set $B_\uniondisjointe$ is the subset of $B$ which is completely joined to $K$ and such that there
	is no edge between $A$ and $B_\uniondisjointe$ and no edge between $I$ and
	$B_\uniondisjointe$.
\item The set $C_\fulljoin$ is the subset of $C$ which is completely joined to $A \cup B$ and to
	$K$ and such that there is no edge between $I$ and $C_\fulljoin$.
\item The set $C_\uniondisjointe$ is the subset of $C$ which is completely joined to $K$ and such that there
	is no edge between $A \cup B$ and $C_\uniondisjointe$ and no edge between
	$I$ and $C_\uniondisjointe$.
\item Each of the sets $B_\fulljoin,B_\uniondisjointe,C_\fulljoin$ and
	$C_\uniondisjointe$ have at least $3t+1$ elements.
\end{itemize}

\begin{figure}[!htb]
	\centering
	\scalebox{.5}{\begin{tikzpicture}

	\node[draw,fill,circle] (A0) at (-3,9) {};
	\node (A0Label) at (-3.6,9) {\LARGE$x$};
	\node[draw,fill,circle] (A1) at (-3,6.5) {};
	\node[draw,fill,circle] (A2) at (-3,4) {};

	\node[draw,fill,circle] (B0) at (3,7) {};
	\node (B0Label) at (2.4,7) {\LARGE$b$};
	\node[draw,fill,circle] (B1) at (3,6) {};
	\node[draw,fill,circle] (B2) at (3,4) {};
	\node[draw,fill,circle] (B3) at (3,3) {};

	\node[draw,fill,circle] (C0) at (0,-3) {};
	\node[draw,fill,circle] (C1) at (0,-4) {};
	\node (C1Label) at (0,-4.6) {\LARGE$c$};
	\node[draw,fill,circle] (C2) at (0,-6.5) {};
	\node[draw,fill,circle] (C3) at (0,-9) {};
	\node[draw,fill,circle] (C4) at (0,-10) {};

	\node[draw,fill,circle] (K0) at (9,1) {};
	\node[draw,fill,circle] (K1) at (9,-1) {};

	\node[draw,fill,circle] (I0) at (-9,2) {};
	\node (I0Label) at (-9.6,2) {\LARGE$y$};
	\node[draw,fill,circle] (I1) at (-9,1) {};
	\node[draw,fill,circle] (I2) at (-9,0) {};
	\node[draw,fill,circle] (I3) at (-9,-1) {};
	\node[draw,fill,circle] (I4) at (-9,-2) {};

	\draw[-] (B2) to (A0);
	\draw[-] (B2) to (A1);
	\draw[-] (B2) to (A2);
	\draw[-] (B3) to (A0);
	\draw[-] (B3) to (A1);
	\draw[-] (B3) to (A2);

	\draw[-] (C0) to (A0);
	\draw[-] (C0) to (A1);
	\draw[-] (C0) to (A2);
	\draw[-] (C0) to (B0);
	\draw[-] (C0) to (B1);
	\draw[-] (C0) to (B2);
	\draw[-] (C0) to (B3);
	\draw[-] (C1) to (A0);
	\draw[-] (C1) to (A1);
	\draw[-] (C1) to (A2);
	\draw[-] (C1) to (B0);
	\draw[-] (C1) to (B1);
	\draw[-] (C1) to (B2);
	\draw[-] (C1) to (B3);

	\draw[-] (K0) to (B0);
	\draw[-] (K0) to (B1);
	\draw[-] (K0) to (B2);
	\draw[-] (K0) to (B3);
	\draw[-] (K0) to (C0);
	\draw[-] (K0) to (C1);
	\draw[-] (K0) to (C2);
	\draw[-] (K1) to (B0);
	\draw[-] (K1) to (B1);
	\draw[-] (K1) to (B2);
	\draw[-] (K1) to (B3);
	\draw[-] (K1) to (C0);
	\draw[-] (K1) to (C1);
	\draw[-] (K1) to (C2);

	\draw[-] (I3) to (K1);
	\draw[-] (I4) to (A2);
	\draw[-] (I2) to (C3);
	\draw[-] (I1) to (A0);
	\draw[-] (I0) to (A0);

	\node[rectangle,
		draw,
		minimum width = 2cm, 
		minimum height = 9cm] (A) at (A1) {};
	\node (Alabel) at (-5,11) {\huge$A$};

	\node[rectangle,
		draw,
		minimum width = 2cm, 
		minimum height = 9cm] (B) at (3,6.5) {};
	\node (Blabel) at (1,11) {\huge$B$};
	\node (BEgalLabel) at (3.5,7.5) {\Large$B_\uniondisjointe$};
	\node (BPlusLabel) at (3.5,4.5) {\Large$B_\fulljoin$};

	\draw[-] (2,8) to (4,8);
	\draw[-] (2,5) to (4,5);

	\node[rectangle,
		draw,
		minimum width = 2cm, 
		minimum height = 9cm] (C) at (0,-6.5) {};
	\node (Clabel) at (-2,-2) {\huge$C$};
	\node (CPlusLabel) at (-0.5,-4.5) {\Large$C_\fulljoin$};
	\node (CEgalLabel) at (-0.5,-7.5) {\Large$C_\uniondisjointe$};

	\draw[-] (-1,-5) to (1,-5);
	\draw[-] (-1,-8) to (1,-8);

	\node[rectangle,
		draw,
		minimum width = 2cm, 
		minimum height = 6cm] (K) at (9,0) {};
	\node (Klabel) at (11,3) {\huge$K$};

	\node[rectangle,
		draw,
		minimum width = 2cm, 
		minimum height = 6cm] (I) at (-9,0) {};
	\node (Ilabel) at (-11,3) {\huge$I$};
\end{tikzpicture}}
	\caption{Structure of a nested $t$-module}
	\label{fig:nestedTModule}
\end{figure}
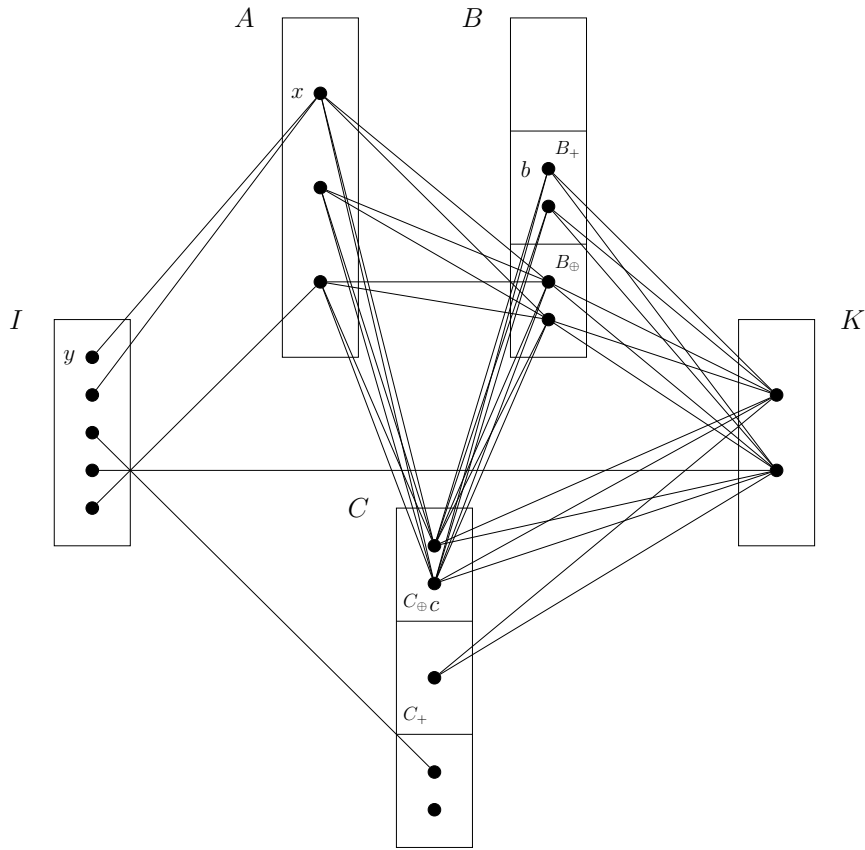

Figure~\ref{fig:nestedTModule} shows a representation of a nested $t$-module. Before stating the reduction
rule, let us observe that if one can provide the sets $A,B,C,K$ and $I$, then the subsets
$B_\fulljoin,C_\fulljoin,B_\uniondisjointe,C_\uniondisjointe$ are polynomial to compute.

\begin{redrule} {\rm (nested $t$-module rule)}\label{tmodulerule}
If $G$ has a nested $t$-module, edit every edge between $A$ and $I$ and every non edge between $A$ and $K$.
\end{redrule}

\begin{lemma} \label{soundness}
The nested $t$-module reduction rule is safe.
\end{lemma}

\begin{proof}
First, observe that if $t=0$ then $A \cup B \cup C$ is a module which is not a comodule. Indeed, since $K
\neq \varnothing$ and $B_\fulljoin \neq \varnothing$ by hypothesis, $A \cup B \cup C$ cannot be a
connected component. Moreover, since $I \neq \varnothing$ and $B_\fulljoin \neq \varnothing$ by hypothesis, $A \cup
B \cup C$ cannot be a connected component in the complement of $G$. Thus, $A \cup B \cup C$ should have
been reduced by Rule~\ref{modulerule} since it is not an independent set as $A \neq \varnothing$ and
$B_\fulljoin \neq \varnothing$ and $A \cap B_\fulljoin = \varnothing$. Now consider the case $t>0$ and assume that
there is an edge $xy$ with $x\in A$ and $y\in I$. Denote by $H$ a minimum cograph edition of $G$ by
$S$ with size at most $k$. By Lemma~\ref{lem:module_budget}, there are at most $t$ pairs of $S$ between
$A$ and $C_\fulljoin\cup B_\uniondisjointe$ ($A$ is a $t$-module of size $\abs{A} > k+t$) and at most $t$ pairs of $S$
between $I$ and $C_\fulljoin\cup B_\uniondisjointe$ ($A \cup B \cup C$ is a $t$-module of size $\abs{A \cup B \cup C} > k +
t$). We denote by $C_\fulljoin'$ (resp $B_\uniondisjointe'$) the subset of $C_\fulljoin$ (resp $B_\uniondisjointe$) which is not incident to one of
these $2t$ pairs. These sets have size at least $t+1$ as $\abs{B_\uniondisjointe} > 3t$ and $\abs{C_\fulljoin} > 3t$. Since
$A \cup B$ is also a $t$-module of size $\abs{A \cup B} > k + t$, not every pair
between $C_\fulljoin'$ and
$B_\uniondisjointe'$ are edited so there exists an edge $cb$ with $c\in
C_\fulljoin'$ and $b\in B_\uniondisjointe'$ such
that $cb\notin S$. A representation of these vertices on a nested $t$-module can be found on
Figure~\ref{fig:nestedTModule}. In particular, the only pair of vertices inside $\{c,b,x,y\}$ which can
be in $S$ is $xy$. Since $yxcb$ is an induced $P_4$, the pair $xy$ must belong to $S$. The same
argument holds for any non edge between $A$ and $K$.
\end{proof}

It is not clear that one can check if the nested $t$-module rule applies in polynomial time. However,
it suffices to be able to correctly guess the sets $A,B,C,K$, and $I$.

Now that we have stated our four reduction rules, let us describe how our kernel works on input $G$.

\begin{enumerate}
	\item Apply these four reduction rules in any order until none is applicable. This gives us a graph $G'$.
	\item If $k$ is small (less than $559$ as we will see in Corollary~\ref{cor:kernel}), do a brute force
to check whether $G'$ can be made a cograph with less than $k$ edge editions.
	\item If $\abs{V(G')}$ is less than some bound in $k$ (a $\grando{}{k^2 \log k}$ that will be given in
Corollary~\ref{cor:kernel}), return $G'$. If not, return any negative instance of size less than $k^2 \log k$ of the cograph $k$-editing
problem (which is answering \ofg{no}).
\end{enumerate}

As we will see later, this algorithm runs in polynomial time in $n = \abs{V(G)}$ so it is
a kernel of size $\grando{}{k^2 \log k}$ for the cograph $k$-edge editing problem.

\newpage
\section{The combinatorial lemma}

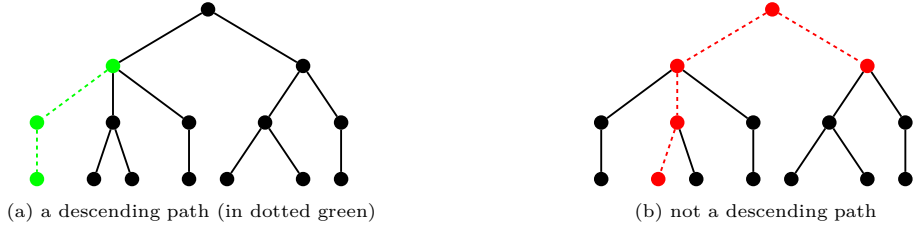
\begin{figure}[t]
	\centering
	\begin{subfigure}{0.45\textwidth}
		\centering
		\scalebox{0.5}{\begin{tikzpicture}[line width=0.5mm]
	\node[draw,fill,circle]{} [sibling distance = 5cm]
		child{node[draw,fill,green,circle]{} [sibling distance = 2cm]
			child{node[draw,fill,green,circle]{} [dashed,green]
				child{node[draw,fill,circle,green,solid]{}}
			}
			child{node[draw,fill,circle]{} [sibling distance = 1cm]
				child{node[draw,fill,circle]{}}
				child{node[draw,fill,circle]{}}
			}
			child{node[draw,fill,circle]{}
				child{node[draw,fill,circle]{}}
			}
		}
		child{node[draw,fill,circle]{} [sibling distance = 2cm]
			child{node[draw,fill,circle]{}
				child{node[draw,fill,circle]{}}
				child{node[draw,fill,circle]{}}
			}
			child{node[draw,fill,circle]{}
				child{node[draw,fill,circle]{}}
			}
		};
\end{tikzpicture}}
		\caption{a descending path (in dotted green)}
	\end{subfigure}
	\hfill
	\begin{subfigure}{0.45\textwidth}
		\centering
		\scalebox{0.5}{\begin{tikzpicture}[line width=0.5mm]
	\node[draw,fill,circle,red]{} [sibling distance = 5cm]
		child{node[draw,fill,red,circle]{} [sibling distance = 2cm, dashed,red]
			child{node[draw,fill,circle,black,solid]{} [solid,black]
				child{node[draw,fill,circle,solid]{}}
			}
			child{node[draw,fill,circle,red,solid]{} [sibling distance = 1cm]
				child{node[draw,fill,circle,red,solid]{}}
				child{node[draw,fill,circle,black,solid]{} [solid,black]}
			}
			child{node[draw,fill,circle,black,solid]{} [solid,black]
				child{node[draw,fill,circle,solid]{}} [solid,black]
			}
		}
		child{node[draw,fill,red,circle]{} [sibling distance = 2cm, dashed,red]
			child{node[draw,fill,circle,black,solid]{} [solid,black]
				child{node[draw,fill,circle,solid]{}}
				child{node[draw,fill,circle,solid]{}}
			}
			child{node[draw,fill,circle,black,solid]{} [solid,black]
				child{node[draw,fill,circle,solid]{}}
			}
		};
\end{tikzpicture}}
		\caption{not a descending path}
	\end{subfigure}
	\caption{A descending path in green and a non descending path (in dotted
	red)}
	\label{fig:descending_path_and_counter_example}
\end{figure}

In a rooted tree (or forest), a path which starts from a node and ends in one of its descendants is a
\emph{descending path} (see Figure~\ref{fig:descending_path_and_counter_example}). We assume here that
$T$ is a rooted tree or a forest which is edge-covered by a collection $\mathcal P$ consisting of $k$
descending paths $P_1,\dots ,P_k$. We do not assume that $\mathcal P$ is minimum, and there could be
some multiple copies of the same path. Given some constant $c \geq 1$, we say that a descending path
$Q$ which is a subpath of some $P_i$ with at least one edge is \emph{$c$-sparse} if it intersects at
most $\abs{E(Q)}/c$ paths of $\mathcal P$ on at least one edge. We start by giving a sufficient condition
for $T$ to have a $c$-sparse path in $\mathcal P$ in the particular case where $T$ is a path. This will
be useful for our proof of Lemma~\ref{lem:foret}.

\begin{lemma} \label{pathcase}
	Let $T$ be a rooted tree which is a path and $\mathcal P$ be a set of $k$ (descending) paths that
	covers all the edges of $T$. If $\abs{E(T)} \geq 4ck$ then there exists a $c$-sparse path $Q$.
\end{lemma}

\begin{proof}
Consider a minimum cover $\mathcal C$ of $T$ by some paths of $\mathcal P$. Free to reorder the paths,
we assume that $\mathcal C$ is the set $P_1,\dots ,P_r$ and that the starting point of $P_i$ is an
ascendant of the starting point of $P_j$ when $1\leq i<j\leq r$. Note that since $\mathcal C$ is a
minimum cover, $P_i$ is disjoint from $P_j$ whenever $1\leq i<j-1\leq r$. Now we partition $\mathcal C$
into ${\mathcal C}_o$ (paths with odd indices) and ${\mathcal C}_e$ (paths with even indices). Without
loss of generality, we assume that the sum of the numbers of edges of the paths in ${\mathcal C}_o$ is more than
$2ck$. We will show that some path $P_i \in \mathcal P$ is $c$-sparse.

Assume by contradiction that every path $P_i \in {\mathcal C}_o$ is not $c$-sparse, and thus intersects
$d_i$ paths of $\mathcal P$ with $d_i > \abs{E(P_i)} / c$. By the fact that ${\mathcal C}$ is a minimum
cover, no path in ${\mathcal P}$ intersects more than two paths in ${\mathcal C}_o$. Since the paths of
$\mathcal{C}_o$ are disjoint, the total number of paths in ${\mathcal P}$ intersecting a path of
${\mathcal C}_o$ is more than

\centers{$\Sum{i=1}{r} \f{d_i}{2} > \Sum{i=1}{r} \f{\abs{E(P_i)}}{2c} \geq k$}

\noindent which is a contradiction.
\end{proof}

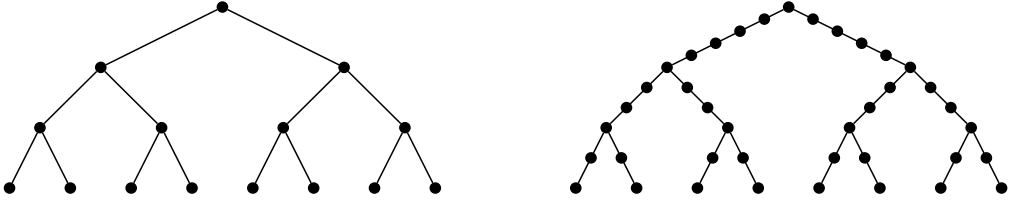
\begin{figure}[t]
	\centering
	\begin{subfigure}{0.45\textwidth}
		\centering
		\scalebox{0.4}{\begin{tikzpicture}[line width=0.5mm]
	\node[draw,fill,circle] (0) at (0,6) {};
	\node[draw,fill,circle] (1) at (-4,4) {};
	\node[draw,fill,circle] (2) at (4,4) {};
	\node[draw,fill,circle] (3) at (-6,2) {};
	\node[draw,fill,circle] (4) at (-2,2) {};
	\node[draw,fill,circle] (5) at (2,2) {};
	\node[draw,fill,circle] (6) at (6,2) {};
	\node[draw,fill,circle] (7) at (-7,0) {};
	\node[draw,fill,circle] (8) at (-5,0) {};
	\node[draw,fill,circle] (9) at (-3,0) {};
	\node[draw,fill,circle] (10) at (-1,0) {};
	\node[draw,fill,circle] (11) at (1,0) {};
	\node[draw,fill,circle] (12) at (3,0) {};
	\node[draw,fill,circle] (13) at (5,0) {};
	\node[draw,fill,circle] (14) at (7,0) {};

	\draw[-] (0) to (1);
	\draw[-] (0) to (2);
	\draw[-] (1) to (3);
	\draw[-] (1) to (4);
	\draw[-] (3) to (7);
	\draw[-] (3) to (8);
	\draw[-] (4) to (9);
	\draw[-] (4) to (10);
	\draw[-] (2) to (5);
	\draw[-] (2) to (6);
	\draw[-] (5) to (11);
	\draw[-] (5) to (12);
	\draw[-] (6) to (13);
	\draw[-] (6) to (14);
\end{tikzpicture}}
	\end{subfigure}
	\hfill
	\begin{subfigure}{0.45\textwidth}
		\centering
		\scalebox{0.4}{\usetikzlibrary{calc}
\begin{tikzpicture}[line width=0.5mm]
	\node[draw,fill,circle] (0) at (0,6) {};
	\node[draw,fill,circle] (1) at (-4,4) {};
	\node[draw,fill,circle] (2) at (4,4) {};
	\node[draw,fill,circle] (3) at (-6,2) {};
	\node[draw,fill,circle] (4) at (-2,2) {};
	\node[draw,fill,circle] (5) at (2,2) {};
	\node[draw,fill,circle] (6) at (6,2) {};
	\node[draw,fill,circle] (7) at (-7,0) {};
	\node[draw,fill,circle] (8) at (-5,0) {};
	\node[draw,fill,circle] (9) at (-3,0) {};
	\node[draw,fill,circle] (10) at (-1,0) {};
	\node[draw,fill,circle] (11) at (1,0) {};
	\node[draw,fill,circle] (12) at (3,0) {};
	\node[draw,fill,circle] (13) at (5,0) {};
	\node[draw,fill,circle] (14) at (7,0) {};

	\node[draw,fill,circle] at ($(0) !1/5! (1)$) {};
	\node[draw,fill,circle] at ($(0) !2/5! (1)$) {};
	\node[draw,fill,circle] at ($(0) !3/5! (1)$) {};
	\node[draw,fill,circle] at ($(0) !4/5! (1)$) {};

	\node[draw,fill,circle] at ($(0) !1/5! (2)$) {};
	\node[draw,fill,circle] at ($(0) !2/5! (2)$) {};
	\node[draw,fill,circle] at ($(0) !3/5! (2)$) {};
	\node[draw,fill,circle] at ($(0) !4/5! (2)$) {};

	\node[draw,fill,circle] at ($(1) !1/3! (3)$) {};
	\node[draw,fill,circle] at ($(1) !2/3! (3)$) {};

	\node[draw,fill,circle] at ($(1) !1/3! (4)$) {};
	\node[draw,fill,circle] at ($(1) !2/3! (4)$) {};

	\node[draw,fill,circle] at ($(2) !1/3! (5)$) {};
	\node[draw,fill,circle] at ($(2) !2/3! (5)$) {};

	\node[draw,fill,circle] at ($(2) !1/3! (6)$) {};
	\node[draw,fill,circle] at ($(2) !2/3! (6)$) {};

	\node[draw,fill,circle] at ($(3) !1/2! (7)$) {};
	\node[draw,fill,circle] at ($(3) !1/2! (8)$) {};
	\node[draw,fill,circle] at ($(4) !1/2! (9)$) {};
	\node[draw,fill,circle] at ($(4) !1/2! (10)$) {};
	\node[draw,fill,circle] at ($(5) !1/2! (11)$) {};
	\node[draw,fill,circle] at ($(5) !1/2! (12)$) {};
	\node[draw,fill,circle] at ($(6) !1/2! (13)$) {};
	\node[draw,fill,circle] at ($(6) !1/2! (14)$) {};

	\draw[-] (0) to (1);
	\draw[-] (0) to (2);
	\draw[-] (1) to (3);
	\draw[-] (1) to (4);
	\draw[-] (2) to (5);
	\draw[-] (2) to (6);
	\draw[-] (3) to (7);
	\draw[-] (3) to (8);
	\draw[-] (4) to (9);
	\draw[-] (4) to (10);
	\draw[-] (5) to (11);
	\draw[-] (5) to (12);
	\draw[-] (6) to (13);
	\draw[-] (6) to (14);
\end{tikzpicture}}
	\end{subfigure}
	\caption{Example of a subdivided tree}
	\label{fig:tree_subdivise}
\end{figure}

If Lemma~\ref{pathcase} would be true for trees, we could derive a quadratic kernel for cograph
edge edition. Unfortunately the following tree provides a counter example: consider a balanced binary
tree with $k$ leaves where $k$ is a power of 2. Now subdivide the two top edges $k/2$ times,
the four next edges $k/4$ times, etc. In the end, the edge connected to the leaves are subdivided once.
Figure~\ref{fig:tree_subdivise} illustrates this procedure for $k=8$.

\begin{figure}[t]
	\centering
	\begin{subfigure}{.45\textwidth}
		\centering
		\scalebox{.4}{\begin{tikzpicture}[line width=0.5mm]
	\node[draw,fill,circle] (0) at (0,6) {};
	\node (rLabel) at (0,6.5) {\LARGE$r$};
	\node[draw,fill,circle] (1) at (-4,4) {};
	\node[draw,fill,red,circle] (2) at (4,4) {};
	\node (u1Label) at (4.5,4) {\LARGE$u_1$};
	\node[draw,fill,circle] (3) at (-6,2) {};
	\node[draw,fill,circle] (4) at (-2,2) {};
	\node[draw,fill,red,circle] (5) at (2,2) {};
	\node[draw,fill,circle] (6) at (6,2) {};
	\node[draw,fill,circle] (7) at (-7,0) {};
	\node[draw,fill,circle] (8) at (-5,0) {};
	\node[draw,fill,circle] (9) at (-3,0) {};
	\node[draw,fill,circle] (10) at (-1,0) {};
	\node[draw,fill,circle] (11) at (1,0) {};
	\node[draw,fill,red,circle] (12) at (3,0) {};
	\node (xLabel) at (3,-0.5) {\LARGE$x$};
	\node[draw,fill,circle] (13) at (5,0) {};
	\node[draw,fill,circle] (14) at (7,0) {};

	\node[draw,fill,circle] at ($(0) !1/5! (1)$) {};
	\node[draw,fill,circle] at ($(0) !2/5! (1)$) {};
	\node[draw,fill,circle] at ($(0) !3/5! (1)$) {};
	\node[draw,fill,circle] at ($(0) !4/5! (1)$) {};

	\node[draw,fill,circle] (02) at ($(0) !1/5! (2)$) {};
	\node[draw,fill,red,circle] (002) at ($(0) !2/5! (2)$) {};
	\node (u0Label) at (1.6,5.7) {\LARGE$u_0$};
	\node[draw,fill,red,circle] (0002) at ($(0) !3/5! (2)$) {};
	\node[draw,fill,red,circle] (00002) at ($(0) !4/5! (2)$) {};

	\node[draw,fill,circle] at ($(1) !1/3! (3)$) {};
	\node[draw,fill,circle] at ($(1) !2/3! (3)$) {};

	\node[draw,fill,circle] at ($(1) !1/3! (4)$) {};
	\node[draw,fill,circle] at ($(1) !2/3! (4)$) {};

	\node[draw,fill,red,circle] at ($(2) !1/3! (5)$) {};
	\node[draw,fill,red,circle] at ($(2) !2/3! (5)$) {};

	\node[draw,fill,circle] at ($(2) !1/3! (6)$) {};
	\node[draw,fill,circle] at ($(2) !2/3! (6)$) {};

	\node[draw,fill,circle] at ($(3) !1/2! (7)$) {};
	\node[draw,fill,circle] at ($(3) !1/2! (8)$) {};
	\node[draw,fill,circle] at ($(4) !1/2! (9)$) {};
	\node[draw,fill,circle] at ($(4) !1/2! (10)$) {};
	\node[draw,fill,circle] at ($(5) !1/2! (11)$) {};
	\node[draw,fill,red,circle] at ($(5) !1/2! (12)$) {};
	\node[draw,fill,circle] at ($(6) !1/2! (13)$) {};
	\node[draw,fill,circle] at ($(6) !1/2! (14)$) {};

	\draw[-] (0) to (1);
	\draw[-] (0) to (02) to (002);
	\draw[-,dashed,red] (002) to (0002) to (00002) to (2);
	\draw[-] (1) to (3);
	\draw[-] (1) to (4);
	\draw[-,dashed,red] (2) to (5);
	\draw[-] (2) to (6);
	\draw[-] (3) to (7);
	\draw[-] (3) to (8);
	\draw[-] (4) to (9);
	\draw[-] (4) to (10);
	\draw[-] (5) to (11);
	\draw[-,dashed,red] (5) to (12);
	\draw[-] (6) to (13);
	\draw[-] (6) to (14);
\end{tikzpicture}}
	\end{subfigure}
	\hfill
	\begin{subfigure}{.45\textwidth}
		\centering
		\scalebox{.4}{\begin{tikzpicture}[line width=0.5mm]
	\node[draw,fill,circle] (0) at (0,6) {};
	\node (rLabel) at (0,6.5) {\LARGE$r$};
	\node[draw,fill,circle] (1) at (-4,4) {};
	\node[draw,fill,red,circle] (2) at (4,4) {};
	\node (u1Label) at (5.25,4) {\LARGE$u_0 = u_1$};
	\node[draw,fill,circle] (3) at (-6,2) {};
	\node[draw,fill,circle] (4) at (-2,2) {};
	\node[draw,fill,red,circle] (5) at (2,2) {};
	\node[draw,fill,circle] (6) at (6,2) {};
	\node[draw,fill,circle] (7) at (-7,0) {};
	\node[draw,fill,circle] (8) at (-5,0) {};
	\node[draw,fill,circle] (9) at (-3,0) {};
	\node[draw,fill,circle] (10) at (-1,0) {};
	\node[draw,fill,circle] (11) at (1,0) {};
	\node[draw,fill,red,circle] (12) at (3,0) {};
	\node (xLabel) at (3,-0.5) {\LARGE$x$};
	\node[draw,fill,circle] (13) at (5,0) {};
	\node[draw,fill,circle] (14) at (7,0) {};

	\node[draw,fill,circle] at ($(0) !1/5! (1)$) {};
	\node[draw,fill,circle] at ($(0) !2/5! (1)$) {};
	\node[draw,fill,circle] at ($(0) !3/5! (1)$) {};
	\node[draw,fill,circle] at ($(0) !4/5! (1)$) {};

	\node[draw,fill,circle] (02) at ($(0) !1/5! (2)$) {};
	\node[draw,fill,circle] (002) at ($(0) !2/5! (2)$) {};
	\node[draw,fill,circle] (0002) at ($(0) !3/5! (2)$) {};
	\node[draw,fill,circle] (00002) at ($(0) !4/5! (2)$) {};

	\node[draw,fill,circle] at ($(1) !1/3! (3)$) {};
	\node[draw,fill,circle] at ($(1) !2/3! (3)$) {};

	\node[draw,fill,circle] at ($(1) !1/3! (4)$) {};
	\node[draw,fill,circle] at ($(1) !2/3! (4)$) {};

	\node[draw,fill,red,circle] at ($(2) !1/3! (5)$) {};
	\node[draw,fill,red,circle] at ($(2) !2/3! (5)$) {};

	\node[draw,fill,circle] at ($(2) !1/3! (6)$) {};
	\node[draw,fill,circle] at ($(2) !2/3! (6)$) {};

	\node[draw,fill,circle] at ($(3) !1/2! (7)$) {};
	\node[draw,fill,circle] at ($(3) !1/2! (8)$) {};
	\node[draw,fill,circle] at ($(4) !1/2! (9)$) {};
	\node[draw,fill,circle] at ($(4) !1/2! (10)$) {};
	\node[draw,fill,circle] at ($(5) !1/2! (11)$) {};
	\node[draw,fill,red,circle] at ($(5) !1/2! (12)$) {};
	\node[draw,fill,circle] at ($(6) !1/2! (13)$) {};
	\node[draw,fill,circle] at ($(6) !1/2! (14)$) {};

	\draw[-] (0) to (1);
	\draw[-] (0) to (02) to (002);
	\draw[-] (002) to (0002) to (00002) to (2);
	\draw[-] (1) to (3);
	\draw[-] (1) to (4);
	\draw[-,dashed,red] (2) to (5);
	\draw[-] (2) to (6);
	\draw[-] (3) to (7);
	\draw[-] (3) to (8);
	\draw[-] (4) to (9);
	\draw[-] (4) to (10);
	\draw[-] (5) to (11);
	\draw[-,dashed,red] (5) to (12);
	\draw[-] (6) to (13);
	\draw[-] (6) to (14);
\end{tikzpicture}}
	\end{subfigure}
	\caption{some non $3$-sparse paths}
	\label{fig:tree_not_3_sparse}
\end{figure}
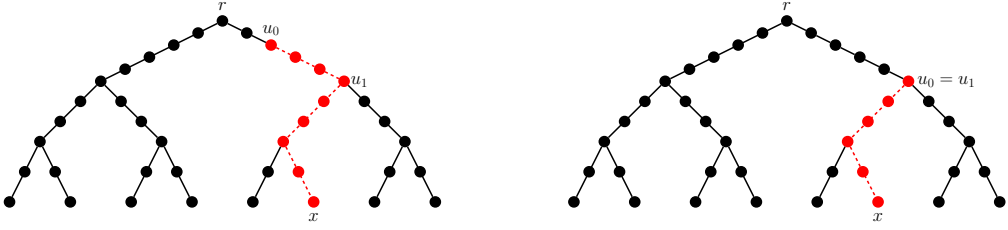

The family ${\mathcal P}$ consists of all the $k$
root-leaf paths. The total size of the tree $T$ is\footnote{More precisely, the tree has $k \log_2(k) + 2k
- 1$ nodes.} $\Omega(k \log_2 k)$. Let us prove that $T$ does not
contain any $3$-sparse path $Q$. By contradiction, assume that there exists a $3$-sparse path $Q$. By
definition, $Q$ is a subpath of some element of $\mathcal P$ hence it is a descending path. Denote by
$u_0$ the first node of $Q$, by $x$ its last node and by $u_1$ its first node of degree $3$ or $r$ if
$r \in V(Q)$. For $u,u'$
two nodes of $Q$, we denote by $Q[u,u']$ the subpath of $Q$ delimited by $u$ and $u'$. Finally, let $d$ be the number of paths of $\mathcal P$ that $Q$
intersects on at least one edge. Figure~\ref{fig:tree_not_3_sparse} helps to understand the following
counts. We have,

\centers{$\abs{E(Q)} = \abs{E(Q[u_0,u_1])} + \abs{E(Q[u_1,x])}$}

\begin{itemize}
	\item If $u_0 = u_1$, then
\newline
\noindent
\begin{tabular*}{\linewidth}{@{\extracolsep{\fill}} l l @{\extracolsep{0pt} } c
	@{ \extracolsep{0pt}} l @{\extracolsep{\fill}} r}
			& $\abs{E(Q)}$ & $=$ & $\abs{E(Q[u_1,x])}$ &\\
			& & $=$ & $\log_2(2d) + 1 + \f{2d}{2} + \f{2d}{4} + \cdots +
			\f{2d}{2^{\log_2(2d)}} - 1$ &\\
			& & $=$ & $\log_2(2d) + 2d - 1$ &\\
			& & $=$ & $\log_2(d) + 2d$
\end{tabular*}

	\noindent Hence, $\abs{E(Q)} < 3d$ which is a contradiction.

	\item If $u_0 \neq u_1$, then
		\centers{$\abs{E(Q[u_1,x])} = \log_2(d) + d - 1$}

	\noindent (replace $2d$ by $d$ in the last count) and moreover,
	$\abs{E(Q[u_0,u_1])} \leq d$. Hence,
		\centers{$\abs{E(Q)} \leq d + \log_2(d) + d - 1 < 3d$}

	\noindent which is a contradiction.
\end{itemize}

This proves that Lemma~\ref{pathcase} is not true anymore when $T$ is not a path
as we just provided a counter example for $c = 3$. Hence it seems that an extra
$\log_2 k$ factor is needed for trees, and we indeed show that it suffices.

\begin{lemma}
	\label{lem:foret}
Let $T$ be a forest and $\mathcal P$ be a set of $k$ descending paths that covers all edges of $T$. If
$\abs{E(T)} \geq 4ck (1 + \log_2 k)$ then there exists a $c$-sparse path $Q$.
\end{lemma}

\begin{proof}
We proceed by induction on $k$. The case $k=1$ is clear since any subpath of $P_1$ with at least $c$
edges is $c$-sparse. If $T$ is a forest, say $T$ is composed by the trees $T_1,\ldots,T_r$ with $r \geq
2$, we define
for all $1 \leq i \leq r$, the set $\mathcal{P}_i$ of the paths of $\mathcal{P}$ whose vertices belong
to $T_i$ and we denote by $k_i$ the size of $\mathcal{P}_i$. Let us show that there exists $T_i$ such that $\abs{E(T_i)} \geq 4c k_i (1 +
\log_2 k_i)$. Assume by contradiction that for all $1 \leq i \leq r$, $\abs{E(T_i)} < 4c k_i (1 + \log_2 k_i)$. Then

\centers{$\abs{E(T)} = \Sum{i=1}{r} \abs{E(T_i)} < 4c \Sum{i=1}{r} k_i (1 + \log_2 k_i) = 4ck + 4c \,
\Sum{i=1}{r} k_i \log_2 k_i$}

\noindent Since $x \mapsto x \log_2 x$ is convex on $\intff{1}{k}$,

\centers{$\forall x \in \intff{1}{k} \qquad x \log_2 x \leq \f{k \log_2(k)}{k-1}(x-1)$}

\noindent
\begin{tabular*}{\linewidth}{@{\extracolsep{\fill}} l l @{\extracolsep{0pt} } c
	@{ \extracolsep{0pt}} l @{\extracolsep{\fill}} r}
	hence & $\Sum{i=1}{r} k_i \log_2 k_i$ & $\leq$ & $\f{k \log_2 k}{k-1} \,
	\Sum{i=1}{r} (k_i - 1)$&\\
	& & $\leq$ & $k \, \log_2(k) \, \f{k-r}{k-1}$&\\
	& & $<$ & $k \, \log_2 k$ & (since $r \geq 2$)
\end{tabular*}

\noindent which leads to the contradiction $\abs{E(T)} < 4c k (1 + \log_2 k)$. We can now apply the induction
hypothesis to $T_i$. In what follows, we assume $T$ to be a (rooted) tree.

Let us construct a descending path $R$ of $\mathcal P$ which starts at the root
$r$ of $T$ and such that for every node $u \in R$, the child $v$ of $u$ whose
subtree intersects the maximum number of paths of $\mathcal P$ is in $R$. In other words, $R$
follows the subtree that intersects the maximum number of paths of $\mathcal P$.
If $R$ has at least $4ck$ edges, we conclude by Lemma~\ref{pathcase}. If not, we
remove from $T$ every edge and every vertex of $R$ and for any node $u$ of $R$ and any $v$ child of
$u$ in $T$ not in $R$, we add a new vertex $v'$ and add the edge $v'v$. We obtain a forest $F$ as
illustrated on Figure~\ref{fig:tree2forest}. Observe that we can identify any
new edge $v'v$ with the old edge $uv$ and thus every connected component in the new forest is edge
covered by $\mathcal P$.

\begin{figure}[t]
	\centering
	\begin{subfigure}{0.45\textwidth}
		\centering
		\scalebox{.45}{\begin{tikzpicture}[line width=0.5mm]
	\node[draw,fill,circle] (0) at (0,4) {};
	\node[draw,fill,circle] (1) at (-5,2) {};
	\node[draw,fill,circle] (2) at (0,2) {};
	\node[draw,fill,circle] (3) at (4,2) {};
	\node[draw,fill,circle] (4) at (-6,0) {};
	\node[draw,fill,circle] (5) at (-4,0) {};
	\node[draw,fill,circle] (6) at (-2,0) {};
	\node[draw,fill,circle] (7) at (2,0) {};
	\node[draw,fill,circle] (8) at (3,0) {};
	\node[draw,fill,circle] (9) at (5,0) {};
	\node[draw,fill,circle] (10) at (-7,-2) {};
	\node[draw,fill,circle] (11) at (-6,-2) {};
	\node[draw,fill,circle] (12) at (-5,-2) {};
	\node[draw,fill,circle] (13) at (-4,-2) {};
	\node[draw,fill,circle] (14) at (-3,-2) {};
	\node[draw,fill,circle] (15) at (-2,-2) {};
	\node[draw,fill,circle] (16) at (-1,-2) {};
	\node[draw,fill,circle] (17) at (2,-2) {};
	\node[draw,fill,circle] (18) at (3,-2) {};
	\node[draw,fill,circle] (19) at (4,-2) {};
	\node[draw,fill,circle] (20) at (6,-2) {};

	\draw[-] (0) to (1);
	\draw[-,dashed,red] (0) to (2);
	\draw[-] (0) to (3);

	\draw[-] (1) to (4);
	\draw[-] (1) to (5);

	\draw[-] (4) to (10);
	\draw[-] (4) to (11);
	\draw[-] (4) to (12);

	\draw[-,dashed,red] (2) to (6);
	\draw[-] (2) to (7);

	\draw[-] (6) to (13);
	\draw[-,dashed,red] (6) to (14);
	\draw[-] (6) to (15);
	\draw[-] (6) to (16);

	\draw[-] (7) to (17);
	\draw[-] (7) to (18);

	\draw[-] (3) to (8);
	\draw[-] (3) to (9);

	\draw[-] (9) to (19);
	\draw[-] (9) to (20);
\end{tikzpicture}}
	\end{subfigure}
	\hfill
	\begin{subfigure}{0.45\textwidth}
		\centering
		\scalebox{0.45}{\begin{tikzpicture}[line width=0.5mm]
	\node[draw,fill,circle] (0_1) at (-.5,4) {};
	\node[draw,fill,circle] (0_2) at (.5,4) {};
	\node[draw,fill,circle] (1) at (-5,2) {};
	\node[draw,fill,circle] (2) at (0,2) {};
	\node[draw,fill,circle] (3) at (4,2) {};
	\node[draw,fill,circle] (4) at (-6,0) {};
	\node[draw,fill,circle] (5) at (-4,0) {};
	\node[draw,fill,circle] (6_1) at (-2.5,0) {};
	\node[draw,fill,circle] (6_2) at (-2,0) {};
	\node[draw,fill,circle] (6_3) at (-1.5,0) {};
	\node[draw,fill,circle] (7) at (2,0) {};
	\node[draw,fill,circle] (8) at (3,0) {};
	\node[draw,fill,circle] (9) at (5,0) {};
	\node[draw,fill,circle] (10) at (-7,-2) {};
	\node[draw,fill,circle] (11) at (-6,-2) {};
	\node[draw,fill,circle] (12) at (-5,-2) {};
	\node[draw,fill,circle] (13) at (-4,-2) {};
	\node[draw,fill,circle] (15) at (-2,-2) {};
	\node[draw,fill,circle] (16) at (-1,-2) {};
	\node[draw,fill,circle] (17) at (2,-2) {};
	\node[draw,fill,circle] (18) at (3,-2) {};
	\node[draw,fill,circle] (19) at (4,-2) {};
	\node[draw,fill,circle] (20) at (6,-2) {};

	\draw[-] (0_1) to (1);
	\draw[-] (0_2) to (3);

	\draw[-] (1) to (4);
	\draw[-] (1) to (5);

	\draw[-] (2) to (7);
	\draw[-] (4) to (10);
	\draw[-] (4) to (11);
	\draw[-] (4) to (12);

	\draw[-] (6_1) to (13);
	\draw[-] (6_2) to (15);
	\draw[-] (6_3) to (16);

	\draw[-] (7) to (17);
	\draw[-] (7) to (18);

	\draw[-] (3) to (8);
	\draw[-] (3) to (9);

	\draw[-] (9) to (19);
	\draw[-] (9) to (20);
\end{tikzpicture}}
	\end{subfigure}
	\caption{exploding a tree with path $R$ (in dotted red)}
	\label{fig:tree2forest}
\end{figure}
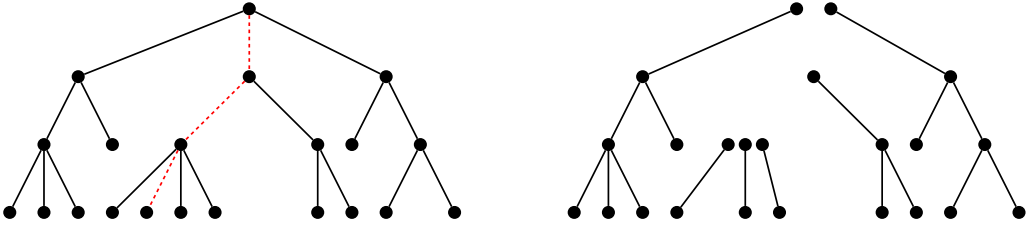

We denote by $C_1,\ldots,C_p$ the connected components obtained. Observe that every component
intersects at most $k/2$ paths of ${\mathcal P}$ by our choice of $R$. For $1 \leq i \leq p$ we denote
by $k_i$ the number of paths of ${\mathcal P}$ intersecting the component $C_i$. There exists $1 \leq i
\leq p$ so that $C_i$ has at least $4ck_i (1 + \log_2 k_i)$ edges (hence we conclude by the induction
hypothesis). Indeed, assume by contradiction that every $C_i$ has strictly less that $4ck_i (1 + \log_2
k_i)$ edges, then the total number of edges in $T$ satisfies

\centers{$\abs{E(T)} < 4ck+\Sum{i=1}{p} 4ck_i (1 + \log_2 k_i) \qquad$ with $\qquad \Sum{i=1}{p} k_i
\leq k \quad$ and $\quad k_i \leq k/2$ for all $i$}

\noindent Since $x \mapsto x \log_2 x$ is convex on $\intff{1}{k/2}$,

\centers{$\forall x \in \intff{1}{k/2} \qquad x \log_2 x \leq \f{k/2 \log_2(k/2)}{k/2 -
1} (x-1)$}

\leftcenters{so}{$\Sum{i=1}{p} k_i \log_2 k_i \leq \f{k/2 \pa{\log_2(k) - 1}}{k/2 - 1}
\pa{\Sum{i=1}{p} k_i - p} \leq \f{k/2 \pa{\log_2(k) - 1}}{k/2 - 1} (k - p)$}

\noindent
\begin{tabular*}{\linewidth}{@{\extracolsep{\fill}} l l @{\extracolsep{0pt} } c
	@{ \extracolsep{0pt}} l @{\extracolsep{\fill}} r}
	then & $4ck + \Sum{i=1}{p} 4c k_i (1 + \log_2 k_i)$ & $\leq$ & $4ck + 4ck + 4c \, \f{k \log_2(k) - k}{k-2}
	(k-p)$ &\\
		  & & $\leq$ & $4ck + 4ck + 4c (k \log_2(k) - k)$ & since $p \geq 2$\\
		  & & $\leq$ & $4ck(1 + \log_2 k)$ &
\end{tabular*}

\noindent Thus $T$ has strictly less than $4ck (1 + \log_2 k)$ edges, a contradiction.
\end{proof}

\section{The \texorpdfstring{$\boldsymbol{k^2 \log k}$}{quasi-quadratic} kernel}

In what follows, $k$ is an integer, $G$ has $n$ vertices and $H$ is a minimum
cograph edit of $G$
by $S$ where $S$ has size at most $k$. Since our goal is to show a
quasi-quadratic vertex kernel, we assume
moreover that $n>k+1$, otherwise we would be done. Moreover, we assume that none of the first three
rules apply to $G$. Our goal is to show that the fourth rule applies if $G$ is large enough (more
than a quasi-quadratic function of $k$). A vertex of $G$ is \emph{edited} if it belongs to some pair in $S$.

We consider $T$ to be the decomposition tree of the cograph $H$. If $u$ is a node of $T$, the set of
descendants of $u$ which are leaves is denoted by $De(u)$. We also see it as a set of vertices of $G$.
We now define a particular subtree $T'$ of $T$ induced by the nodes $u$ such that $\abs{De(u)} \geq k+2$. It
is indeed a subtree since we have $De(u) \subseteq De(parent(u))$ for all node $u$ which is not the
root of $T$.

\begin{lemma} \label{sizeT'}
The tree $T'$ has at least $n/(k+1)-2k$ nodes.
\end{lemma}

\begin{proof}
	For $u \in V(T')$, we denote by $A_{T'}(u)$ the set of $x \in De(u)$ such that the path from $u$ to $x$
	in $T$ does not contain any node of $T'$ except $u$. In other words, for a vertex $x \in V(G)$, we
	have that $x \in A_{T'}(u)$ if and only
	if $u$ is the closest ancestor of $x$ in $T$ that belongs to $T'$. Define

	\centers{$L(T') := \{u \in V(T') \ | \ A_{T'}(u) \neq \varnothing\}$}

	For $u \in V(T')$, define $Be(u)$ (resp $Bn(u)$) to be the set of children $v$ of $u$ in $T$ such
	that $v \notin V(T')$ and $De(v)$ contains (resp does not contain) an edited vertex.
	Figure~\ref{fig:schema_lemme_size_tprime} illustrates this setting. Observe that

	\centers{$V(G) = \Union{u \in L(T')}{} \pa{\pa{\Union{v \in Be(u)}{} De(v)} \cup \pa{\Union{v \in
	Bn(u)}{} De(v)}}$}

	\noindent Indeed, take $x \in V(G)$ and consider its closest ancestor $u$ in $T$ which belongs to
	$T'$. This is well defined since $V(T') \ni r$ as $n \geq k+2$. Let $v$ be the child of $u$ on this
	path (we could have $v=x$). By definition of $u$, we have that $v \notin V(T')$ and either $v \in
	Be(u)$ or $v \in Bn(u)$. In both case, $x \in De(v)$ so

	\centers{$V(G) \subseteq \Union{u \in L(T')}{} \pa{\pa{\Union{v \in Be(u)}{} De(v)} \cup \pa{\Union{v \in
	Bn(u)}{} De(v)}}$}

	\begin{figure}[t]
		\centering
		\scalebox{.6}{\usetikzlibrary{shapes.geometric}
\begin{tikzpicture}
	\node[ellipse,
		draw = black,
		minimum width = 6cm, 
		minimum height = 1cm] (e1) at (-3,1) {$B_n(u)$};

	\node[ellipse,
		draw = black,
		minimum width = 2cm, 
		minimum height = 1cm] (e2) at (3,1) {$\qquad\qquad Be(u)$};

	\node[ellipse,
		draw = black,
		minimum width = 14cm, 
		minimum height = 3.5cm] (e4) at (-1,-2) {$De(u)$};

	\draw[dashed,very thick,red] plot [smooth, tension=1] coordinates {(3,-3) (9,-2) (13,1)};
	\node[draw,fill,circle] (0) at (5,5) {};
	\node[draw,fill,circle] (1) at (-1,3) {};
	\node (uLabel) at (-1.5,3) {$u$};
	\node[draw,fill,circle] (2) at (11,3) {};
	\node[draw,fill,circle] (3) at (-5,1) {};
	\node[draw,fill,circle] (4) at (-1,1) {};
	\node[draw,fill,circle] (5) at (3,1) {};
	\node[draw,fill,circle] (6) at (9,1) {};
	\node[draw,fill,circle] (7) at (13,1) {};
	\node[draw,fill,circle] (8) at (-6,-1) {};
	\node[draw,fill,circle] (9) at (-5,-1) {};
	\node[draw,fill,circle] (10) at (-4,-1) {};
	\node[draw,fill,circle] (11) at (-2,-1) {};
	\node[draw,fill,circle] (12) at (0,-1) {};
	\node[draw,fill,circle] (13) at (2,-1) {};
	\node[draw,fill,circle] (14) at (4,-1) {};
	\node[draw,fill,circle] (15) at (1,-3) {};
	\node[draw,fill,circle] (16) at (3,-3) {};

	\draw[-] (0) to (1);
	\draw[-] (0) to (2);

	\draw[-] (1) to (3);
	\draw[-] (1) to (4);
	\draw[-] (1) to (5);

	\draw[-] (2) to (6);
	\draw[-] (2) to (7);

	\draw[-] (3) to (8);
	\draw[-] (3) to (9);
	\draw[-] (3) to (10);

	\draw[-] (4) to (11);
	\draw[-] (4) to (12);

	\draw[-] (5) to (13);
	\draw[-] (5) to (14);

	\draw[-] (13) to (15);
	\draw[-] (13) to (16);
\end{tikzpicture}}
		\caption{Structure of $T'$ with an edited edge in dotted red}
		\label{fig:schema_lemme_size_tprime}
	\end{figure}
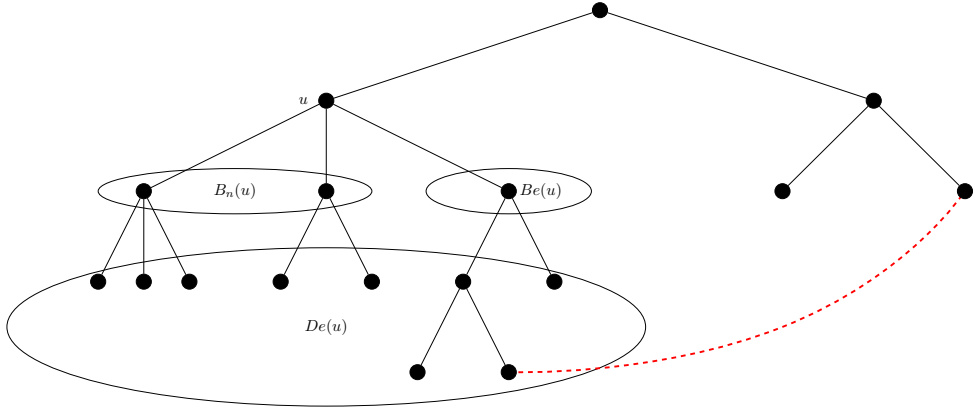

	Observe that since there are at most $2k$ edited vertices and since for all $u \in L(T')$ and all
	$v \in Be(u)$, $\abs{De(v)} \leq k+1$, we have that

	\centers{$\abs{\Union{u \in L(T')}{} \, \Union{v \in Be(u)}{} De(v)} \leq 2k(k+1)$}

	\leftcenters{Hence}{$\abs{\Union{u \in L(T')}{} \, \Union{v \in Bn(u)}{} De(v)} \geq n - 2k(k+1)$}

	\noindent Observe that the sets involved in the union on $u \in L(T')$ are pairwise disjoint. Indeed, for all $x \in V(G)$ there exists a
unique $u \in L(T')$ such that $x \in A_{T'}(u)$. So,

	\centers{$\Sum{u \in L(T')}{} \abs{\Union{v \in Bn(u)}{} De(v)} \geq n - 2k(k+1)$}

	\noindent Moreover, for all $u \in L(T')$, the set $\Union{v \in Bn(u)}{} De(v)$ is a module as it
	does not contain any edited vertex. Moreover, it is not a comodule as it would have been removed by
	Rule~\ref{componentrule}. Hence, by Rule~\ref{eliminationrule}, its size is at most $k+1$.
	Hence,

	\centers{$\abs{V(T')} \geq \abs{L(T')} \geq \f{n - 2k(k+1)}{k+1} =
	\f{n}{k+1} - 2k$}
\end{proof}

The edited pairs $xy$ in $S$ can be analyzed with respect to the tree $T$. In particular, every pair
$xy$ in $S$ corresponds to the path $P_{xy}$ of $T$ which connects the leaves $x$ and $y$. If we denote
by $z$ the least common ancestor of $x$ and $y$ in $T$, we obtain two descending paths $P_{zx}$ and $P_{zy}$ which
form an edge-partition of $P_{xy}$. There are at most $2k$ such descending paths in $T$ called
\emph{edit paths}. 

\begin{lemma} \label{editioncoverT'}
Every edge of $T'$ belongs to an edit path, except possibly $k$ edges incident to the root of $T'$.
\end{lemma}

\begin{proof}
	Let $u \in V(T')$ and assume that $u$ is not the root of $T'$. Let $p$ be its parent node. Assume that
	the edge $up$ does not belong to any edit path. Then $De(u)$ is a module. By definition of
	$T'$, $\abs{De(u)} \geq k+2 > k+1$ so $De(u)$ must be a comodule by Rules~\ref{eliminationrule} and~\ref{modulerule}. Hence, $p$
	is the root of $T'$. This proves that every edge of $T'$ not incident to its root belongs to an
	edit path. Moreover, $De(u)$ must contain an edited pair since it would have been
	removed by Rule~\ref{componentrule} otherwise. Hence, $T'$ has at most $k$ edges which does not
	belong to an edit path and all of these edges are incident to its root.
\end{proof}

Let us denote by $T''$ the forest obtained from $T'$ when we remove the edges that does not belong to
an edit path. By definition, $T''$ is edge covered by the edit paths.

\begin{theorem} \label{kernel}
If $T''$ has a $51$-sparse path with respect to the edit paths, then the nested $t$-module reduction
rule applies to $G'$. Moreover, one can detect such a nested $t$-module in polynomial time.
\end{theorem}

\begin{proof}
	Let $Q_0$ be a $51$-sparse path with respect to the edit paths. Recall that by definition,

	\begin{itemize}
		\item $Q_0$ is a subpath of some edit path
		\item $Q_0$ intersects (on at least one edge) $\ell$ edit paths
		\item $1 \leq \ell \leq \abs{E(Q_0)}/51$
	\end{itemize}

	\noindent We consider a subpath $Q$ of $Q_0$ with
	$\abs{E(Q)} = 51 \ell$ edges. A node $u$ of $Q$ which is not the first or the last node (and hence has
	a descendant $u'$ in $Q$) is \emph{free} if $De(u) \setminus De(u')$ does not contain any edited
	vertex. In particular, $De(u) \setminus De(u')$ is a module of $G$. In such a case, we denote by $F_Q(u)$ the set $De(u)
	\setminus De(u')$. Note that $F_Q(u)$ is a module which is not a comodule since $u$ is not the root
	of $T$ (it is not the first node of $Q$). Hence, by Rules~\ref{eliminationrule}
	and~\ref{modulerule}, $F_Q(u)$ is an independent set. Let us prove that any non free node $u \in V(Q)$
	satisfies that $u \, parent(u)$ or $uu'$ belong to an edit path where $u'$ is the
	child of $u$ in $Q$. First, if there exists an edited pair $xy$ such that $x \in F_Q(u)$ and $y \in
	(V \setminus De(u)) \cup De(u')$ then

	\begin{itemize}
		\item either $y \in V \setminus De(u)$ in which case $P$ intersects $Q$
			on the edge $u \, parent(u)$
		\item or $y \in De(u')$ and so $P$ intersects $Q$ on $uu'$.
	\end{itemize}

	\noindent In both cases, the edit path $P$ intersects $Q$ hence intersects $Q_0$. Now, let us assume by contradiction
	that for every edited pair $xy$ both $x$ and $y$ belong to $F_Q(u)$. Define

	\centers{$S' := \{xy \in S \ | \ x \notin F_Q(u) \vee y \notin F_Q(u)\}$}

	\noindent and observe that $\abs{S'} < \abs{S}$ as it must exists and edited vertex in $F_Q(u)$.
	Denote by $G'$ the edition of $G$ by $S'$. Let us show that $G'$ is a cograph. Indeed, $G'$
	coincide with $G$ except on $F_Q(u)$. Assume by contradiction that there exists an induced $P_4$ in
	$G'[F_Q(u)]$, say $x_1,x_2,x_3,x_4$. Since $F_Q(u)$ is an independent set in $H$, each of the
	pairs $x_1x_2,x_2x_3$ and $x_3x_4$ belong to $S$ so we could have
	made $G$ a cograph with fewer
	edits by removing $x_1x_2$ from $S$. So $G'$ is a cograph which again contradicts the minimality
	of $S$.

	This implies that $Q$ cannot have more than $2\ell$ non-free nodes since $Q_0$ is $51$-sparse (two
	consecutive non free nodes may correspond to the same intersection).

	Assume now that $u$ and its child $u'$ in $Q$ are both free and that $u$ is
	labelled $\fulljoin$ (thus $u'$
	is labelled $\uniondisjointe$). Since
	$F_Q(u)$ and $F_Q(u')$ are independent sets, all vertices of $F_Q(u')$ are children of $u'$ and all
	vertices of $F_Q(u)$ are children of $u''$, a child of $u$ not in $Q$
	(Figure~\ref{fig:consecutiveFreeNodes} illustrates this situation). Pick a vertex $x \in F_Q(u)$
	and a vertex $x' \in F_Q(u')$. The crucial observation is that $V \setminus De(u)$ is exactly the
	set of vertices $y$ distinct from $x$ and $x'$ such that $\{x,x'\}$ is a module of $G[x,x',y]$. Indeed,

	\begin{itemize}
		\item the vertices of $V \setminus De(u)$ have this property as both $x$ and $x'$ are non
			edited and $u$ is labelled $\fulljoin$,
		\item the vertices in $De(u') \setminus \{x'\}$ are joined to $x$ and not to $x'$
		\item and the vertices of $F_Q(u) \setminus \{x\}$ are joined to $x'$ and not to $x$.
	\end{itemize}

	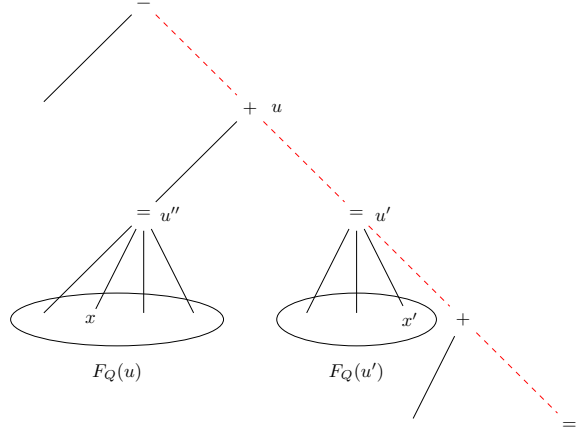
\begin{figure}[t]
		\centering
		\scalebox{.7}{\begin{tikzpicture}
	\node[ellipse,
		draw = black,
		minimum width = 4cm, 
		minimum height = 1cm] (e1) at (-4.5,-2) {};

	\node (e1Label) at (-4.5,-3) {$F_Q(u)$};

	\node[ellipse,
		draw = black,
		minimum width = 3cm, 
		minimum height = 1cm] (e2) at (0,-2) {};

	\node (e2Label) at (0,-3) {$F_Q(u')$};

	\node[circle] (0) at (-4,4) {$=$};
	\node (1) at (-6,2) {};
	\node[circle] (u) at (-2,2) {$+$};
	\node (uLabel) at (-1.5,2) {$u$};
	\node[circle] (uPrime) at (0,0) {$=$};
	\node (uPrimeLabel) at (0.5,0) {$u'$};
	\node[circle] (2) at (2,-2) {$+$};
	\node[circle] (uSecond) at (-4,0) {$=$};
	\node (uSecondLabel) at (-3.5,0) {$u''$};
	\node[circle] (3) at (4,-4) {$=$};

	\node (f0) at (-6,-2) {};
	\node (x) at (-5,-2) {$x$};
	\node (f1) at (-4,-2) {};
	\node (f2) at (-3,-2) {};
	\node (f4) at (-1,-2) {};
	\node (f5) at (0,-2) {};
	\node (xPrime) at (1,-2) {$x'$};
	\node (f7) at (1,-4) {};

	\draw[-,dashed,red] (0) to (u) to (uPrime) to (2) to (3);
	\draw[-] (0) to (1);
	\draw[-] (u) to (uSecond);

	\draw[-] (uSecond) to (f0);
	\draw[-] (uSecond) to (x);
	\draw[-] (uSecond) to (f1);
	\draw[-] (uSecond) to (f2);

	\draw[-] (uPrime) to (f4);
	\draw[-] (uPrime) to (f5);
	\draw[-] (uPrime) to (xPrime);
	\draw[-] (2) to (f7);
\end{tikzpicture}}
		\caption{Consecutive free nodes in the path $Q$ drawn in dotted red}
		\label{fig:consecutiveFreeNodes}
	\end{figure}

	\noindent Hence, if one provides $x$ and $x'$, we can compute $De(u)$ in polynomial time (in $n$).
	In the following, we refer to such a couple $(u,u')$ in $Q$ as a \emph{cut}.

	\begin{claim}
		There exists a cut in any subpath of $Q$ that has at least $8 \ell$ edges.
	\end{claim}

	\begin{proof}
		Let $Q'$ be a subpath of $Q$ that has at least $8 \ell$ edges. It
		suffices to show that there exists three consecutive nodes in $Q'$ that
		are free (either a sequence $\uniondisjointe,\fulljoin,\uniondisjointe$
		or a sequence $\fulljoin,\uniondisjointe,\fulljoin$). Assume by
		contradiction that every sequence of three consecutive nodes in $Q'$
		contains a non free node. Then, the number of intersections between $Q'$
		and some paths of $\mathcal P$ is at least

		\centers{$\f{1}{2} \times \f{\abs{V(Q')}}{3} \geq \f{1}{2} \times \f{8 \ell}{3} = \f{4 \ell}{3} >
		\ell$}

		\noindent which contradicts the fact that $Q_0$ is $51$-sparse.
	\end{proof}

	We now pick three cuts $(u,u')$, $(v,v')$ and $(w,w')$ in $Q$ such that $(u,u')$ is chosen in the
	range $\intn{43 \ell}{50 \ell}$ so among the $9 \ell$ last nodes of $Q$ but not among the $\ell$ last ones, $(v,v')$ are in the middle of $Q$
	(precisely chosen in the range $\intn{23 \ell}{30 \ell}$) and $(w,w')$ are chosen in the first
	$10 \ell$ nodes of $Q$ but not among the $2 \ell$ first ones (in the range $\intn{3 \ell}{10
	\ell}$). Take $x \in F_Q(u)$, $x' \in F_Q(u')$,
	$y \in F_Q(v)$, $y' \in F_Q(v')$, $z \in F_Q(w)$ and $z' \in F_Q(w')$. Define $A := De(u)$, then $B
	:= De(v) \setminus A$ and finally $C := De(w)\setminus (A\cup B)$. The vertices of $V(G) \setminus
	(A \cup B \cup C)$ (which is equal to $V \setminus De(w)$) which are connected to $x$ form the set
	$K$, and the other vertices form the set $I$. Figure~\ref{fig:nestedLModule} illustrates how these
	elements are distributed on the tree $T$.

	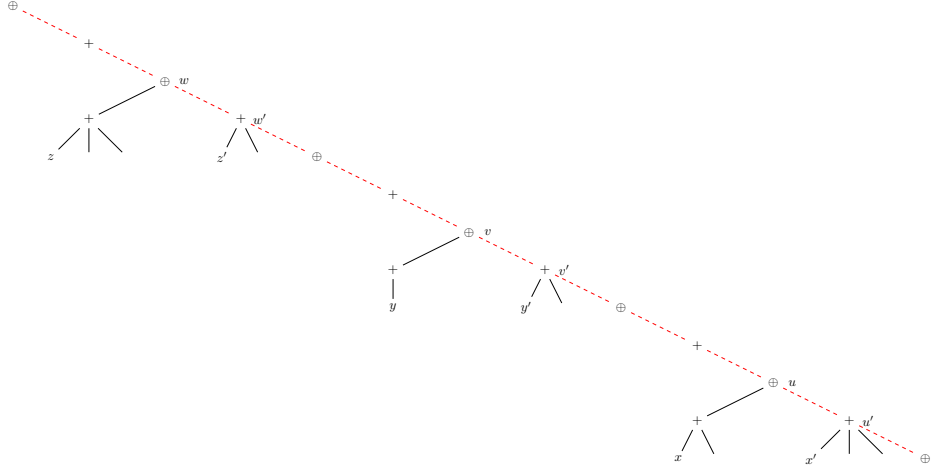
\begin{figure}[t]
		\centering
		\scalebox{.5}{\begin{tikzpicture}
	\node (0) at (-14,7) {$\fulljoin$};
	\node (1) at (-12,6) {$\uniondisjointe$};
	\node (w) at (-10,5) {$\fulljoin$};
	\node (wLabel) at (-9.5,5) {$w$};
	\node (wPrime) at (-8,4) {$\uniondisjointe$};
	\node (wPrimeLabel) at (-7.5,4) {$w'$};
	\node (3) at (-6,3) {$\fulljoin$};
	\node (4) at (-4,2) {$\uniondisjointe$};
	\node (v) at (-2,1) {$\fulljoin$};
	\node (vLabel) at (-1.5,1) {$v$};
	\node (vPrime) at (0,0) {$\uniondisjointe$};
	\node (vPrimeLabel) at (0.5,0) {$v'$};
	\node (5) at (2,-1) {$\fulljoin$};
	\node (6) at (4,-2) {$\uniondisjointe$};
	\node (u) at (6,-3) {$\fulljoin$};
	\node (uLabel) at (6.5,-3) {$u$};
	\node (uPrime) at (8,-4) {$\uniondisjointe$};
	\node (uPrimeLabel) at (8.5,-4) {$u'$};
	\node (7) at (10,-5) {$\fulljoin$};
	\node (8) at (-12,4) {$\uniondisjointe$};
	\node (9) at (-4,0) {$\uniondisjointe$};
	\node (10) at (4,-4) {$\uniondisjointe$};
	\node (z) at (-13,3) {$z$};
	\node (f0) at (-12,3) {};
	\node (f1) at (-11,3) {};
	\node (zPrime) at (-8.5,3) {$z'$};
	\node (f2) at (-7.5,3) {};
	\node (y) at (-4,-1) {$y$};
	\node (yPrime) at (-0.5,-1) {$y'$};
	\node (f3) at (0.5,-1) {};
	\node (x) at (3.5,-5) {$x$};
	\node (f4) at (4.5,-5) {};
	\node (xPrime) at (7,-5) {$x'$};
	\node (f5) at (8,-5) {};
	\node (f6) at (9,-5) {};

	\draw[-,dashed,red] (0) to (1) to (w) to (wPrime) to (3) to (4) to (v) to (vPrime) to (5) to (6) to (u) to
	(uPrime) to (7);

	\draw[-] (w) to (8);
	\draw[-] (v) to (9);
	\draw[-] (u) to (10);

	\draw[-] (8) to (z);
	\draw[-] (8) to (f0);
	\draw[-] (8) to (f1);

	\draw[-] (wPrime) to (zPrime);
	\draw[-] (wPrime) to (f2);

	\draw[-] (9) to (y);

	\draw[-] (vPrime) to (yPrime);
	\draw[-] (vPrime) to (f3);

	\draw[-] (10) to (x);
	\draw[-] (10) to (f4);

	\draw[-] (uPrime) to (xPrime);
	\draw[-] (uPrime) to (f5);
	\draw[-] (uPrime) to (f6);
\end{tikzpicture}}
		\caption{Representation of our nested $\ell$-module (the path $Q$ is in
		dotted red)}
		\label{fig:nestedLModule}
	\end{figure}

	\noindent By construction, these five sets are pairwise
	disjoint. Moreover, $A,B$ and $C$ are nonempty. Let us check that $K \neq \varnothing$ and $I \neq \varnothing$. Since $w$ is not the
	first node of $Q$, it has a parent $p$ and $p$ is labelled $\uniondisjointe$. Consider $t \in De(p) \setminus
	De(w)$ and observe that since $x$ is not edited, $t$ is not connected to $x$. Hence, $I \neq
	\varnothing$. Moreover, $p$ is not the first node of $Q$ (we took $w$ not among the first $2\ell$
	nodes of $Q$) so it has a parent $p'$ and $p'$ is labelled $\fulljoin$. Consider $t' \in De(p') \setminus
	De(p)$ and observe that $t'$ is connected to $x$. Hence, $K \neq \varnothing$.

	Observe that $A$, $A \cup B$ and $A \cup B \cup C$ are $\ell$-modules of size at
	least $k+\ell+1$. Indeed, since $Q$ intersects at most $\ell$ edit paths, there is less than
	$\ell$ edges to edit to make any of them a module. Moreover, let us show that $A$ (hence $A \cup B$ and $A
	\cup B \cup C$) has at least $k + \ell + 1$ elements. Let $\gamma$ be the last node of $Q$. Since
	$\gamma
	\in V(T')$, we have that $\abs{De(\gamma)} > k+1$. Observe that $De(\gamma) \subseteq De(u) = A$
	and that $u$ has at least $\ell$ descendants in $Q$ since we did not take $u$ among the last $\ell$
	nodes of $Q$. So there exists $\ell - 1$ vertices of $G$ in $De(u) \setminus De(\gamma)$. Hence,
	$\abs{A} \geq k+1+\ell$.

	Another important point is that $A,B,C,K$ and $I$ can be constructed if one correctly guesses (in
	time $\grando{}{n^6}$) the six vertices $x,x',y,y',z$ and $z'$. Indeed, we proved that $A$ which is $V
	\setminus De(u)$ is exactly the set of vertices $t$ distinct from $x$ and $x'$ such that $\{x,x'\}$ is a module of
	$G[x,x',t]$. In other words, since both $x$ and $x'$ are not edited,

	\centers{$A = \{a \in V \ | \ ax \in E(G) \wedge ax' \notin E(G)\} \cup \{a \in V \ | \ ax \notin
	E(G) \wedge ax' \in E(G)\}$}

	\noindent Similarly, $B$ and $C$ can be constructed in polynomial time given $y,y',z$ and $z'$.

	Let us define the sets $B_\fulljoin,C_\fulljoin,B_\uniondisjointe$ and
	$C_\uniondisjointe$ as in the definition of nested $t$-module. We denote
	by $U_\fulljoin$ the subset of internal nodes of the subpath $Q[v',u]$ which
	are free and labelled~$\fulljoin$.
	Since $Q[v',u]$ has at least $12\ell$ internal nodes and since there are at most $2\ell$ non free
	nodes, the size of $U_\fulljoin$ is at least $12\ell/2 - 2\ell = 4\ell$.
	Observe that for any $\alpha \in U_\fulljoin$, $F_Q(\alpha)$
	is completely joined to $K$ and to $A$ and there is no edge between $I$ and $F_Q(\alpha)$.
	Moreover, $F_Q(\alpha) \subseteq B$ by definition. Hence,

	\centers{$B_\fulljoin \supseteq \Union{\alpha \in U_\fulljoin}{} F_Q(\alpha)$}

	\noindent which proves that $\abs{B_\fulljoin} \geq 4 \ell > 3 \ell$. We
	prove in a similar manner that the sets $B_\uniondisjointe,C_\fulljoin$ and
	$C_\uniondisjointe$ have size at least $3 \ell + 1$. Recall that we can
	construct $B_\fulljoin,B_\uniondisjointe,C_\fulljoin$ and
	$C_\uniondisjointe$ in polynomial time if we are provided $A,B,C,K$ and $I$.
	Therefore, if indeed $A,B,C,K,I$ is a nested $\ell$-module, we can find it
	in polynomial time.

	In order to show that the $t$-module rule applies, we need to check that there is at least one edge
	or one non edge to edit. In other words, we have to prove that there is either an edge between $A$ and $I$ or a non edge between $A$
	and $K$ in $G$. Since $Q$ is an edit path, there exists $a \in A$ and $s \in V \setminus (A \cup B \cup
	C)$ such that $as \in S$. Since $V \setminus (A \cup B \cup C) = K \cup I$, either $s \in K$ and in
	that case $as$ is a non edge in $G$ or $s \in I$ and $as$ is an edge. In both cases, the
	$t$-module rule applies.
\end{proof}

\begin{coro}
	\label{cor:kernel}
	Cograph editing has a vertex kernel of size $\grando{}{k^2\log k}$.
\end{coro}

\begin{proof}
We assume that we apply the three first rules until none is applicable. We consider the decomposition
tree $T$ of $G$ and the forest $T''$ as previously defined. Recall that $T''$ is obtained from $T'$ by
removing every edge of $T'$ which does not belong to an edit path. By Lemma~\ref{editioncoverT'},
there are at most $k$ such edges. Since the value of $k$ can be supposed
larger than some fixed constant, say $k \geq 559$ here (otherwise we conclude by brute force), we can
assume that the number of edges in $T''$ is at least

\centers{$\f{n}{k+1} - 2k -1 - k \geq \f{409 k^2(1 + \log_2 2k)}{k+1}-3k-1 \geq 408 k (1 + \log_2 2k)$}

\noindent The forest $T''$ is covered by at most $2k$ edit paths, so, by
Lemma~\ref{lem:foret}, it contains a $51$-sparse descending path, and we
conclude by Theorem~\ref{kernel}. If $k \geq 559$ and $n \geq 409 k^2(1 + \log_2
2k)$ and if none of our four rules is applicable, we return any graph of size at
most $409 k^2 (1 + \log_2 2k)$ which cannot be made a cograph with less than $k$
edge editions (this is returning « no »). Hence, we have designed a polynomial
time (in $n$) algorithm that transforms any graph $G_{in}$ into a graph
$G_{out}$ of size at most $409 k^2(1+\log_2 2k)$ such that $G_{in}$ and
$G_{out}$ are equivalent instances of the cograph $k$ editing problem.
\end{proof}

\section{Declaration of interest}

Declarations of interest: none.

\bibliography{bibliographie.bib}

\begin{thebibliography}{10}
\expandafter\ifx\csname url\endcsname\relax
  \def\url#1{\texttt{#1}}\fi
\expandafter\ifx\csname urlprefix\endcsname\relax\def\urlprefix{URL }\fi
\expandafter\ifx\csname href\endcsname\relax
  \def\href#1#2{#2} \def\path#1{#1}\fi

\bibitem{GHPP2013}
S.~Guillemot, F.~Havet, C.~Paul, A.~Perez,
  \href{https://hal.inria.fr/hal-00821612}{{On the (non-)existence of
  polynomial kernels for $P_l$-free edge modification problems}},
  {Algorithmica} 65~(4) (2013) 900--926.
\newline\urlprefix\url{https://hal.inria.fr/hal-00821612}

\bibitem{Man08}
F.~Mancini, Graph modification problems related to graph classes, Ph.D. thesis,
  University of Bergen, Norway (2008).

\bibitem{Natanzon1999ComplexityCO}
A.~Natanzon, R.~Shamir, R.~Sharan, Complexity classification of some edge
  modification problems, in: WG, 1999.

\bibitem{Shamir2002ClusterGM}
R.~Shamir, R.~Sharan, D.~Tsur, Cluster graph modification problems, in: WG,
  2002.

\bibitem{CDF+20}
C.~Crespelle, P.~G. Drange, F.~V. Fomin, P.~A. Golovach,
  \href{https://arxiv.org/abs/2001.06867}{A survey of parameterized algorithms
  and the complexity of edge modification}, CoRR abs/2001.06867 (2020).
\newblock \href {http://arxiv.org/abs/2001.06867} {\path{arXiv:2001.06867}}.
\newline\urlprefix\url{https://arxiv.org/abs/2001.06867}

\bibitem{downey1999parameterized}
R.~Downey, M.~Fellows, M.~Fellows, D.~Gries, F.~Schneider,
  \href{https://books.google.fr/books?id=pt5QAAAAMAAJ}{Parameterized
  Complexity}, Monographs in Computer Science, Springer New York, 1999.
\newline\urlprefix\url{https://books.google.fr/books?id=pt5QAAAAMAAJ}

\bibitem{BodlaenderDFH09}
H.~L. Bodlaender, R.~G. Downey, M.~R. Fellows, D.~Hermelin, On problems without
  polynomial kernels, J.~Comput. Syst. Sci. 75~(8) (2009) 423--434.

\bibitem{CaoC12}
Y.~Cao, J.~Chen, Cluster editing: Kernelization based on edge cuts,
  Algorithmica 64~(1) (2012) 152--169.

\bibitem{ChenM12}
J.~Chen, J.~Meng, A 2k kernel for the cluster editing problem, J.~Comput. Syst.
  Sci. 78~(1) (2012) 211--220.

\bibitem{DBLP:journals/corr/abs-2105-09566}
G.~Bathie, N.~Bousquet, T.~Pierron,
  \href{https://arxiv.org/abs/2105.09566}{(sub)linear kernels for edge
  modification problems towards structured graph classes}, CoRR abs/2105.09566
  (2021).
\newblock \href {http://arxiv.org/abs/2105.09566} {\path{arXiv:2105.09566}}.
\newline\urlprefix\url{https://arxiv.org/abs/2105.09566}

\bibitem{CORNEIL1981163}
D.~Corneil, H.~Lerchs, L.~Burlingham,
  \href{https://www.sciencedirect.com/science/article/pii/0166218X81900135}{Complement
  reducible graphs}, Discrete Applied Mathematics 3~(3) (1981) 163--174.
\newblock \href {https://doi.org/https://doi.org/10.1016/0166-218X(81)90013-5}
  {\path{doi:https://doi.org/10.1016/0166-218X(81)90013-5}}.
\newline\urlprefix\url{https://www.sciencedirect.com/science/article/pii/0166218X81900135}

\end{thebibliography}
\bibliographystyle{elsarticle-num.bst}

\newpage
\appendix

\section*{Link with the rules by Guillemot et al}

We reproduce here the three first reduction rules given by Guillemot et al
in~\cite{GHPP2013} for completeness.

\begin{redrule}\label{RG1}
	Remove the connected components of $G$ which are cographs.
\end{redrule}

\begin{redrule}\label{RG2}
	If $C = G_1 \oplus G_2$ is a connected component of $G$, then replace $C$ by
	$G_1 + G_2$.
\end{redrule}

\begin{redrule}\label{RG3}
	If $M$ is a non-trivial module of $G$ which is strictly contained in a
	connected component and is not an independent set of size at most $k+1$,
	then return the graph $G' + G[M]$ where $G'$ is obtained from $G$ by
	deleting $M$ and adding an independent set of size $\Min{}\{\abs{M}, k+1\}$
	having the same neighborhood than $M$.
\end{redrule}

We will see that our three first rules are equivalent to the three rules of
Guillemot et al in the following sens:

\begin{prop}\label{prop:rulesequivalence}
	A graph $G$ is reduced for the rules of Guillemot et al if and only if it is
	reduced for our three first reduction rules.
\end{prop}

\begin{proof}
	Let $G$ be a reduced instance under Guillemot's rules. Let us show by
	contradiction that none of our three first rule apply. First, we show that
	Rule~\ref{componentrule} does not apply. Assume by contradiction that $G$
	has a comodule $M$ which induces a cograph.

	\begin{itemize}
		\item If $M$ is a connected component, then it should have been removed
			by Rule~\ref{RG1}.
		\item If not, then $M$ is a connected component in the complement of $G$
			and so, in particular, $M$ is a non-trivial module. Indeed, $M \neq
			G$ since otherwise we are done and moreover, $M \neq \emptyset$ as it is a
			connected component in the complement of $G$. So, $M$ is strictly
			contained in a connected component of $G$.

			\begin{itemize}
				\item If $M$ is an independent set of size $\abs{M} \leq k+1$ then
					Rule~\ref{RG3} applies.
				\item Otherwise, let $C$ be the connected component of $G$ in
					which $M$ is included. Observe that $C = M \oplus (C
					\setminus M)$ hence Rule~\ref{RG2} applies.
			\end{itemize}
	\end{itemize}

	Let us now show that Rule~\ref{eliminationrule} does not apply. Assume by
	contradiction that $G$ has a module $M$ so that $\abs{M} > k+1$ inducing an
	independent set.

	\begin{itemize}
		\item If no edge goes out of $M$, then Rule~\ref{RG1} applies as there are
			$\abs{M}$ connected components (which are isolated vertices) that are
			cographs.
		\item Otherwise, take $C$ the connected component that contains $M$
			(which is well defined as $M$ is a module) and observe again that $C
			= M \oplus (C \setminus M)$ hence Rule~\ref{RG2} applies.
	\end{itemize}

	Let us show that Rule~\ref{modulerule} does not apply. Assume by
	contradiction that there exists a module $M$ that is not a comodule and such
	that $G[M]$ contains an edge. Since it is not a comodule, there is an edge
	going out of $M$ and so, since $M$ is a module, it is included (strictly)
	into a connected component. It is non-trivial and it is not an independent set as
	$G[M]$ contains an edge. Hence Rule~\ref{RG3} applies.

	Conversely, let $G$ be a reduced instance under our three first reduction
	rules (Rules~\ref{componentrule},~\ref{eliminationrule} and
	\ref{modulerule}). Let us show by contradiction that none of Guillemot's
	rules (Rules~\ref{RG1}, \ref{RG2} and~\ref{RG3}) apply. First, we show that
	Rule~\ref{RG1} does not apply. Assume by
	contradiction that $G$ has a connected component $M$ which is a cograph. It
	is a comodule so our Rule~\ref{componentrule} applies.

	Let us show that Rule~\ref{RG2} does not apply. Assume by contradiction that there
	exists a connected component $C$ of $G$ such that $C = G_1 \oplus G_2$.

	\begin{itemize}
		\item If $C$ induces a cograph then Rule~\ref{componentrule} applies.
		\item If not, then $C$ has an induced $P_4$. Observe that because of the
			full join, such $P_4$ cannot overlap $G_1$ nor $G_2$. Without loss
			of generality, we assume this $P_4$ to be included in $G_1$. Then,
			$G_1$ is a module that is not a comodule. Indeed, for every $x,y \in
			V(G_1)$, $x$ and $y$ have the same neighborhood in $G_2$ because of
			the full join and no neighbors outside of $C$. Moreover, $G_1$
			cannot be a comodule: it is not a connected component and, whenever
			$C \neq G$, it cannot be a connected component in the complement of
			$G$. It contains an edge so Rule~\ref{modulerule} applies.
	\end{itemize}

	Finally, we show that Rule~\ref{RG3} does not apply. Assume by contradiction that
	there exists a non-trivial module $M$ which is strictly contained in a
	connected component $C$ and is not an independent set of size at most $k+1$.

	\begin{itemize}
		\item If $M$ induces an independent set of size $\abs{M} > k+1$ then
			Rule~\ref{eliminationrule} applies.
		\item Otherwise, $G[M]$ contains an edge. If $M$ is a comodule then, $C
			= M \oplus (C \setminus M)$ so Rule~\ref{RG2} applies which we just proved
			to be impossible. Hence, $M$ is not a comodule, $G[M]$ contains an
			edge and so Rule~\ref{modulerule} applies.
	\end{itemize}
\end{proof}

\end{document}